\documentclass[12p]{iopart}
\usepackage{graphicx}
\usepackage{iopams}
\usepackage{hyperref}
\usepackage[english]{babel}

\begin{document}
\title[A density-responsive scalar-field for singularity regularization and dark energy]{A density-responsive scalar-field framework for singularity regularization and dynamical dark energy}

\author{Martin Drobczyk$^1$}

\address{$^1$Institute of Space Systems, German Aerospace Center, Bremen, Germany}
\ead{martin.drobczyk@dlr.de}

\begin{abstract}
We present a covariant scalar-field framework that unifies the space-time singularity regularization with dynamical dark energy. The theory extends general relativity by introducing a scalar field $\Phi$ whose potential couples to the Lorentz-invariant quantity $X \equiv u_{\alpha} u_{\beta} T^{\alpha\beta}_{\mathrm{matter}}$, ensuring manifest covariance. The resulting density-responsive scalar energy $\rho_\Phi$ exhibits dual behavior: (i) in high-density regimes, it saturates at $\rho_\Phi \leq AM_P^4/2$, providing a Planck-scale upper bound on the total energy density that regularizes classical singularities; (ii) in low-density regimes, it approaches a constant $\rho_\Phi \to AM_U^4$, driving cosmic acceleration as dynamical dark energy. A natural renormalization group evolution with an anomalous dimension $\gamma \approx 0.501$ connects the Planck scale to the meV dark energy scale without fine-tuning. The model makes distinctive, testable predictions: $w_0 \approx -0.99$ and $w_a \approx +0.03$, where the positive $w_a$ distinguishes it from $\Lambda$CDM and standard quintessence models. Despite the novel interaction terms, the fifth forces are suppressed by $\beta_{\rm eff} \propto 1/\rho_m^2$, yielding factors below $10^{-58}$ in laboratory environments, and ensuring compatibility with all precision gravity tests. This framework demonstrates how a single quantum field theory mechanism can simultaneously address UV singularities and IR dark energy, providing concrete predictions for future Stage-IV cosmological surveys.

\end{abstract}

\noindent{\it Keywords}: scalar field, dynamical dark energy, singularity regularization, renormalization group, cosmology

\maketitle


\section{Introduction}
\label{sec:intro}

The reconciliation of general relativity (GR) and quantum mechanics remains a significant challenge in modern physics. This schism manifests in two of the greatest puzzles of cosmology. At the highest energies, GR predicts its own breakdown in the form of space-time singularities \cite{Hawking1973,Penrose1965}, while at the lowest energies observations reveal a universe undergoing accelerated expansion \cite{Perlmutter1999,Planck2018}, driven by a mysterious dark energy component. The leading candidate, a cosmological constant $\Lambda$, suffers from a severe fine-tuning problem: its observed value is approximately 120 orders of magnitude smaller than naive theoretical expectations \cite{Weinberg1989,Padmanabhan2003}. 

These two phenomena, one marking the breakdown of physics at the Planck scale and the other dominating the universe's evolution at the meV scale, may seem disconnected. However, their shared origin in the interplay between energy and space-time geometry has inspired numerous approaches that seek a unified origin.

The quest for quantum gravity has produced several distinct research programs. String theory posits new fundamental entities and extra dimensions, while loop quantum cosmology (LQC) utilizes a quantum description of geometry to resolve the Big Bang singularity via a "bounce" \cite{AshtekarSingh2011}. Another prominent avenue is the asymptotic safety scenario, which proposes that the ultraviolet behaviour of gravity is controlled by a non-trivial fixed point of the renormalization group (RG) flow, rendering the theory predictive and non-perturbatively renormalizable \cite{BonannoKochPlatania2017,Platania2023_Review}. Such RG-based approaches have led to "quantum-improved" black hole models, in which singularities are regularized by the scale-dependent running of gravitational couplings \cite{Bonanno2017_BH}. Complementary to these efforts, the emergent gravity paradigm suggests that the field equations themselves, along with a positive cosmological constant, arise naturally from the statistical mechanics of spacetime's microscopic degrees of freedom \cite{Padmanabhan2021}.

A parallel effort seeks to understand gravity through the language of gauge theories, analogous to the Standard Model, for example, by attempting to derive gravity from first principles and postulating new fundamental symmetries \cite{Partanen2024}.

In addition to these fundamental approaches, complementary self-tuning scenarios have been proposed to address the dark-energy problem. Relaxion frameworks achieve dynamic selection of vacuum energy via an axion-like field interacting with a dark gauge sector \cite{Graham2015,Choi2016}. Sequestering reformulates the cosmological constant problem as a global constraint and effectively decouple vacuum-energy contributions from gravitational dynamics \cite{Kaloper2014}. Screening models, such as chameleon and symmetron theories, provide density-dependent scalar masses that reconcile long-range fifth forces with local gravity tests while reproducing dark-energy behaviour on cosmological scales \cite{Khoury2004,Brax2004,HinterbichlerKhoury2010}. While a quantitative comparison is beyond the scope of this work, we emphasize that unlike our density-dependent potential, these mechanisms typically rely on non-minimal couplings or global constraints.

In contrast to these approaches, we present a phenomenologically driven framework that acts as an effective field-theory extension of GR. Instead of replacing GR or postulating new symmetries, we augment the theory with a single real scalar field $\Phi$, whose potential $U(\Phi,X)$ couples to the manifestly covariant Lorentz scalar
\begin{equation}
  X \equiv u_\alpha u_\beta T_{\rm matter}^{\alpha\beta}.
\label{eq:lorentz_scalar}
\end{equation}
This construction, motivated by Coleman–Weinberg mechanisms in a dense medium \cite{ColemanWeinberg1973,Birrell1982}, allows the field to dynamically adapt to the local energy density.

The equilibrium energy density of this field $\rho_\Phi$, dubbed the \emph{density-responsive scalar energy}, gives rise to a dual-purpose mechanism governed by a characteristic mass-scale $M_U$ that runs with renormalization scale $\mu$. This RG evolution, driven by a natural anomalous dimension $\gamma \approx 0.501$, connects the Planck scale to the meV scale and yields a natural emergence of the observed dark-energy scale once the IR RG scale is tied to a geometric scalar (e.g.\ $\mu=H$ or $\mu=\sqrt{|R|}$), without parameter fine-tuning at the effective field theory (EFT) level (see Section~\ref{subsec:rg_to_cosmo}). From this single underlying principle the framework yields two key predictions:
\begin{enumerate}
  \item \textbf{singularity regularization:} in the high-density ultraviolet regime, $\rho_\Phi$ is capped, bounding the total energy density, sourcing gravity and regularizing classical curvature singularities.
  \item \textbf{dynamical dark energy:} in the low-density infrared regime, $\rho_\Phi$ drives cosmic acceleration with a distinctive, testable equation-of-state evolution $w_a > 0$, setting it apart from $\Lambda$CDM and most standard quintessence models.
\end{enumerate}

The remainder of this paper is organized as follows. In Section~\ref{sec:framework}, we establish a covariant theoretical framework. Section~\ref{sec:rg_flow} details the RG mechanism for generating a scale hierarchy. Section~\ref{sec:phenomenology} explores the physical consequences from singularity regularization to testable cosmological predictions. We conclude and compare to other models in section~\ref{sec:conclusion}.
\section{The covariant scalar field framework}
\label{sec:framework}

The central challenge in constructing a theory in which a scalar field responds to its environment is to define the concept of the ``local energy density'' in a coordinate independent manner. A naive approach might utilize the time-component of the stress-energy tensor $\rho_m = T^0_0$, but this is an inherently frame-dependent quantity. Here, we resolve this foundational issue by constructing a theory using manifestly covariant objects, thereby ensuring its validity under general coordinate transformations.

\subsection{Action and the covariant potential}
\label{subsec:action_potential}

We extend general relativity by introducing a single real scalar field $\Phi$, whose dynamics are governed by the total action
\begin{equation}
  S = \int \mathrm{d}^4x \sqrt{-g} \left[ \frac{M_P^2}{2}R - \frac{1}{2}M_K^2 g^{\mu\nu}\partial_\mu\Phi\partial_\nu\Phi - U(\Phi, X) + \mathcal{L}_{\rm matter} \right].
  \label{eq:action_covariant}
\end{equation}
Here, $M_P = (8\pi G)^{-1/2}$ is the reduced Planck mass,\footnote{Throughout this study, we work with the reduced Planck mass $M_P \equiv (8\pi G)^{-1/2} \simeq 2.435\times 10^{18}\,$GeV, 
not to be confused with the (unreduced) Planck mass $M_{\rm Pl} \equiv G^{-1/2} \simeq 1.220\times 10^{19}\,$GeV.} $R$ is the Ricci scalar, $M_K$ is the kinetic mass scale (typically of order $M_P$), and $\mathcal{L}_{\rm matter}$ is the Lagrangian density for all conventional matter and radiation fields, which are assumed to be minimally coupled to gravity.

The key innovation of our framework is the scalar potential $U$, which is designed to couple the dynamics of $\Phi$ to the surrounding matter-energy environment. To ensure that this coupling is physically meaningful in any reference frame, the potential depends not on a frame-specific density, but on the manifestly covariant Lorentz scalar defined by
\begin{equation}
  X \equiv u_\alpha u_\beta T_{\rm matter}^{\alpha\beta}.
  \label{eq:X_definition}
\end{equation}
In this definition, $T_{\rm matter}^{\alpha\beta}$ is the stress-energy tensor of all matter species, obtained from the matter Lagrangian alone:
\begin{equation}
  T_{\rm matter}^{\mu\nu} \equiv -\frac{2}{\sqrt{-g}}\,
  \frac{\delta\!\left(\sqrt{-g}\,\mathcal{L}_{\rm matter}\right)}{\delta g_{\mu\nu}}.
  \label{eq:Tmn_matter_def}
\end{equation}
The four-velocity $u^\mu$ is uniquely defined as the future-directed, timelike eigenvector of the total stress tensor $T^\mu{}_\nu \equiv (T_{\rm matter}+T_\Phi)^{\mu\alpha} g_{\alpha\nu}$, satisfying
\begin{equation}
  T^\mu{}_\nu\,u^\nu = -\varepsilon\,u^\mu, \qquad u^\mu u_\mu=-1.
  \label{eq:landau_frame}
\end{equation}
This definition corresponds to the Landau (energy) frame \cite{LandauLifshitz1987}, ensuring that the energy flux vanishes: $q^\mu \equiv -\Delta^\mu{}_{\alpha} T^{\alpha\beta} u_\beta = 0$ with $\Delta^{\mu\nu}\equiv g^{\mu\nu}+u^\mu u^\nu$. For multi-fluid or mildly non-equilibrium configurations where $T^\mu{}_\nu$ is diagonalizable with a dominant timelike eigenvector, this prescription yields a unique $u^\mu$. In the cosmological FLRW background relevant for our analysis, $u^\mu$ reduces to the familiar comoving four-velocity $(1,0,0,0)$, and consequently $X=\rho_m$.

The functional form of the potential $U(\Phi, X)$ is motivated by effective field theory considerations, specifically by Coleman--Weinberg-type radiative corrections in a dense background medium, and is designed to exhibit an inverse-tracking behavior \cite{ColemanWeinberg1973, Copeland1998, Birrell1982}:
\begin{equation}
  U(\Phi, X) = M_U^4(\mu) \left[ \frac{A}{1 + X/M_U^4(\mu)} + \frac{h}{2}\left(\Phi - \Phi_{\rm eq}(X)\right)^2 \right].
  \label{eq:potential_form}
\end{equation}
The potential is characterized by three key components: First, $M_U(\mu)$ is the running characteristic energy scale of the potential, which connects different physical regimes via renormalization group evolution, as detailed in Section~\ref{sec:rg_flow}. Second, the leading term governed by the dimensionless constant
\[
  A_{\mathrm{th}}\equiv\frac{1}{64\pi^{2}}\simeq 1.6\times10^{-3},
\]
a canonical one-loop factor, establishes the inverse relationship between the potential's energy and the environmental density $X$. Third, the quadratic term, with a dimensionless curvature $h > 0$, ensures that the potential is stable and possesses a well-defined minimum at an $X$-dependent equilibrium value $\Phi_{\rm eq}(X)$. While this potential form is theoretically well-motivated, we demonstrate in ~\ref{app:potential_robustness} that the core results of our framework—singularity regularization and the RG-driven hierarchy—are robust against variations of this specific functional form.

The scalar field contributes its own stress-energy tensor, obtained by varying the scalar sector with respect to the metric:
\begin{equation}
  T_{\Phi}^{\mu\nu} = M_K^2\,\partial^{\mu}\Phi\,\partial^{\nu}\Phi
  - g^{\mu\nu}\!\left(\frac{1}{2}M_K^2\,\partial_\alpha\Phi\,\partial^\alpha\Phi+U\right)
  - 2\frac{\partial U}{\partial X}\frac{\partial X}{\partial g_{\mu\nu}}.
  \label{eq:Tphi_general}
\end{equation}
The last term encodes the interaction between the scalar and matter sectors through the metric dependence of $X$. Because $U$ depends on $X$, the total stress-energy $T^{\mu\nu}=T_{\rm matter}^{\mu\nu}+T_{\Phi}^{\mu\nu}$ is conserved by the Bianchi identity, but the individual components exchange energy-momentum: $\nabla_\mu T_{\rm matter}^{\mu\nu}=-\nabla_\mu T_{\Phi}^{\mu\nu}$ whenever $\partial U/\partial X \neq 0$. This exchange is the microscopic origin of the fifth-force effects analyzed in Section~\ref{sec:local_gravity_tests}. In the quasi-static regime relevant for cosmological backgrounds (Section~\ref{subsec:eos_from_T}), the kinetic terms become negligible and $T_\Phi^{\mu\nu}$ reduces to the algebraic form given in equation~(\ref{eq:T_phi_equilibrium}).

\subsection{Equilibrium dynamics and the density-responsive energy}
\label{subsec:equilibrium_dynamics}

A crucial feature of the model, which enables a powerful simplification, is the \textit{rapid relaxation} of the field $\Phi$ to this equilibrium value $\Phi_{\rm eq}(X)$. The stability of this minimum is determined by the effective physical mass of the field's fluctuations. Assuming the kinetic scale $M_K$ and potential scale $M_U$ share a common origin and thus run proportionally ($M_K \propto M_U$), the squared mass is given by
\begin{equation}
    m_\Phi^2 = \frac{1}{M_K^2}\frac{\partial^2 U}{\partial \Phi^2}\bigg|_{\Phi_{\rm eq}} = \frac{h M_U^4(\mu)}{M_K^2}.
    \label{eq:phi_mass}
\end{equation}
A natural assumption within this framework is that the kinetic scale $M_K$ and the potential scale $M_U$ share a common physical origin. This implies they should run proportionally under the RG flow, i.e., $M_K(\mu) \propto M_U(\mu)$. While theoretically well-motivated, we analyze the robustness of our framework against violations of this assumption in ~\ref{app:mk_mu_relation}. Under this assumption, the physical mass of the field fluctuations scales as $m_\Phi \propto M_U(\mu)$. 

A detailed stability analysis, provided in ~\ref{app:quasi_static}, confirms that this scaling leads to a relaxation time, $\tau_{\rm roll} \sim m_\Phi^{-1}$, that is always many orders of magnitude shorter than the relevant dynamical timescale (e.g., $H^{-1}$). This vast separation of scales robustly justifies the quasi-static approximation throughout this work, where the field $\Phi$ is always considered to be in its instantaneous minimum, $\Phi \approx \Phi_{\rm eq}(X)$. In this limit, the kinetic energy of the field, $(\partial\Phi)^2$, is negligible compared to its potential energy. Thus, the field's contribution to the total energy budget of the universe becomes a purely algebraic function of the local matter-energy environment $X$ and the running scale $M_U(\mu)$.

We define this fundamental equilibrium energy density as the {\bf density-responsive scalar energy}, denoted $\rho_\Phi$
\begin{equation}
    \rho_\Phi(X, \mu) \equiv U(\Phi_{\rm eq}(X), X; \mu) = \frac{A M_U^8(\mu)}{X + M_U^4(\mu)}.
    \label{eq:rho_phi_definition}
\end{equation}
The central equation encapsulates the core dynamic principle of the model. It exhibits two limiting behaviors that are essential for its dual role in cosmology and singularity regularization
\begin{itemize}
    \item For $X \gg M_U^4$ (high-density environments, e.g., the early universe or black hole interiors), the scalar energy is suppressed: $\rho_\Phi \approx A M_U^8/X$.
    \item For $X \ll M_U^4$ (low-density environments, e.g., the vacuum of late-time cosmology), the scalar energy approaches a constant value: $\rho_\Phi \approx A M_U^4$, thus capable of acting as a source of cosmic acceleration.
\end{itemize}

\paragraph{Validity of the quasi-static approximation.}
The quasi-static limit $\Phi \simeq \Phi_{\rm eq}(X)$ requires that the field relaxes much faster than the background varies, i.e.\ $\tau_{\rm roll}\equiv m_\Phi^{-1}\ll H^{-1}$ and $m_\Phi \gg \left|\dot{X}/X\right|$. These conditions are satisfied by many orders of magnitude across all relevant regimes (cosmic expansion, gravitational collapse, neutron stars). A detailed estimate in~\ref{app:quasi_static} shows that the tracking error is suppressed by factors between $10^{-16}$ and $10^{-60}$, and spatial gradients are negligible on scales $\gg \lambda_\Phi=\hbar c/m_\Phi$. Only near the Planck epoch can the approximation become marginal; there our conclusions rely on the algebraic density bound rather than on a specific EoS.

\subsection{The full stress-energy tensor $T_\Phi^{\mu\nu}$}
\label{subsec:stress_energy_tensor}

To fully characterize the gravitational impact of the scalar field $\Phi$ and analyze potential non-gravitational interactions, we must derive its complete stress-energy tensor. This is obtained by the standard procedure of varying the scalar field part of the action, $S_\Phi$, with respect to the metric $g_{\mu\nu}$:
\begin{equation}
    T_\Phi^{\mu\nu} = -\frac{2}{\sqrt{-g}} \frac{\delta S_\Phi}{\delta g_{\mu\nu}}.
\end{equation}
This variation must account for both the explicit dependence of the action on $g_{\mu\nu}$ and the implicit dependence through Lorentz scalar $X$, which contains $T_{\rm matter}^{\alpha\beta}$. The general result is
\begin{equation}
  \fl T_\Phi^{\mu\nu} = \underbrace{M_K^2 \left( \nabla^\mu\Phi \nabla^\nu\Phi - \frac{1}{2} g^{\mu\nu} \nabla_\lambda\Phi \nabla^\lambda\Phi \right)}_{\mathrm{kinetic\ term}} - g^{\mu\nu}U(\Phi,X) - 2 \frac{\partial U}{\partial X}\frac{\partial X}{\partial g_{\mu\nu}}.
    \label{eq:T_phi_general}
\end{equation}
As established in Section~\ref{subsec:equilibrium_dynamics}, the kinetic term is negligible in quasi-static approximation. The second term represents the contribution of standard potential energy. The third term, however, is new and crucial, as it introduces a direct interaction between the scalar field and matter sector, mediated by the metric.

To evaluate the interaction term, we need $\partial X/\partial g_{\mu\nu}$. Starting from $X = u_\alpha u_\beta T_{\rm matter}^{\alpha\beta}$ and using the fact that $u^\mu$ is the normalized timelike eigenvector of $T_{\rm matter}^{\mu\nu}$, the variation yields (see ~\ref{app:stress_energy_tensor} for the complete derivation)
\begin{equation}
    \frac{\partial X}{\partial g_{\mu\nu}} = -\frac{1}{2}(\rho_m + p_m)u^\mu u^\nu - \frac{1}{2}p_m g^{\mu\nu}.
    \label{eq:dX_dg}
\end{equation}

In the quasi-static limit where $\Phi \approx \Phi_{\rm eq}(X)$, the potential satisfies $U(\Phi_{\rm eq}, X) = \rho_\Phi(X)$. Taking the derivative with respect to $X$
\begin{equation}
    \frac{\partial U}{\partial X}\bigg|_{\Phi_{\rm eq}} = \frac{\partial \rho_\Phi}{\partial X} = -\frac{A M_U^8}{(X + M_U^4)^2} = -\frac{\rho_\Phi}{X + M_U^4}.
\end{equation}

Substituting equations~(\ref{eq:dX_dg}) and this derivative into (\ref{eq:T_phi_general}), we obtain
\begin{equation}
    \eqalign{T_\Phi^{\mu\nu} &\approx -g^{\mu\nu}\rho_\Phi + 2\frac{\rho_\Phi}{X + M_U^4}\left[\frac{1}{2}(\rho_m + p_m)u^\mu u^\nu + \frac{1}{2}p_m g^{\mu\nu}\right] \nonumber \cr
    &= -g^{\mu\nu}\rho_\Phi\left(1 - \frac{p_m}{X + M_U^4}\right) + \frac{(\rho_m + p_m)\rho_\Phi}{X + M_U^4}u^\mu u^\nu \nonumber \cr
    &= \left(-\rho_\Phi + \frac{p_m \rho_\Phi}{X+M_U^4}\right) g^{\mu\nu} + \frac{(\rho_m + p_m)\rho_\Phi}{X+M_U^4} u^\mu u^\nu}
    \label{eq:T_phi_equilibrium}
\end{equation}

The transition to the final form reveals the physical structure: the scalar field contributes  both an isotropic pressure (modified by matter pressure) and an anisotropic stress aligned with the matter flow.

This tensor reveals several crucial physical features beyond those of the conventional scalar-field theories. The $g^{\mu\nu}$ term shows that the effective pressure of the scalar field receives a correction proportional to the local matter pressure $p_m$. More significantly, the $u^\mu u^\nu$ term represents a novel anisotropic stress contribution aligned with the matter flow direction. This term mediates a direct, dynamic energy momentum exchange between the scalar field and the matter sector and constitutes the microscopic origin of the fifth-force effects analyzed in Section~\ref{sec:local_gravity_tests}.

The full dynamics of space-time are now governed by the modified Einstein field equations as follows

\begin{equation}
    G^{\mu\nu} = 8\pi G \left( T_{\rm matter}^{\mu\nu} + T_\Phi^{\mu\nu} \right).
\end{equation}
While the total stress energy is covariantly conserved by the Bianchi identity $\nabla_\mu(T_{\rm matter}^{\mu\nu} + T_\Phi^{\mu\nu}) = 0$, neither component is conserved separately. The non-zero divergence $\nabla_\mu T_\Phi^{\mu\nu} = -\nabla_\mu T_{\rm matter}^{\mu\nu}$ quantifies the energy-momentum transfer. This interaction is the source of potential fifth-force effects and violations of the weak equivalence principle, which we analyze in detail and constrain with observational data in section~\ref{sec:phenomenology}.

\paragraph{Key assumptions and their implications.}
The above derivation relies on two key assumptions that warrant explicit discussion. First, the quasi-static approximation requires $m_\Phi \gg H$, which we have verified holds across all relevant regimes (~\ref{app:quasi_static}). Second, we assumed that the kinetic and potential scales run proportionally under the RG flow: $M_K(\mu) \propto M_U(\mu)$. This assumption, while natural in hidden-sector models, where both scales share a common origin, affects the field's effective mass, $m_\Phi \propto M_U/M_K$. Different running behaviors would modify the regime of validity for our approximations, although the qualitative features of the model are expected to be robust. The complete UV theory determines this relationship from the first principles.

\subsection{Effective equation of state and background cosmology}
\label{subsec:eos_from_T}

Since the scalar sector exchanges energy–momentum with matter ($\nabla_\mu T_\Phi^{\mu\nu}\neq 0$), the separate–conservation shortcut $w=-\frac{1}{3}\,{\rm d}\ln\rho/{\rm d}\ln a$ does not apply. Instead, the effective variables must be read off from the stress tensor. Using equation~(\ref{eq:T_phi_equilibrium}) in FLRW with $u^\mu=(1,0,0,0)$ (where $u^\mu$ denotes the Landau–frame four–velocity defined in Section~\ref{subsec:action_potential}), we define $\rho_\Phi^{\rm eff}\equiv T_{\Phi}^{\mu\nu}u_\mu u_\nu$ and $T_{\Phi}^{ij}=p_\Phi^{\rm eff}g^{ij}$. For late–time dust ($w_m=0$) this yields a closed expression for the instantaneous EoS,
\begin{equation}
  w_\Phi^{\rm eff}(a)= -\frac{\rho_m(a)+M_U^4}{M_U^4+2\,\rho_m(a)}\,,
  \qquad \rho_m(a)=\rho_{m,0}\,a^{-3},
  \label{eq:w_eff_dust_main}
\end{equation}
with the limits $w_\Phi^{\rm eff}\to -1$ for $\rho_m\ll M_U^4$ and $w_\Phi^{\rm eff}\to -\frac{1}{2}$ for $\rho_m\gg M_U^4$. Although the UV limit is finite, the scalar remains negligible at early times because
\begin{equation}
  \frac{\rho_\Phi}{\rho_m}\;\simeq\;A\left(\frac{M_U^4}{\rho_m}\right)^{\!2}\;\propto\;a^{6}\;\to\;0\quad (a\to 0).
\end{equation}
The background Friedmann equation therefore uses the tensor–derived variables,
\begin{equation}
  H^2(a)=\frac{8\pi G}{3}\Big[\rho_m(a)+\rho_r(a)+\rho_\Phi^{\rm eff}(a)\Big].
\end{equation}
If a geometric RG scale is adopted (e.g.\ $\mu=H(a)$ or $\sqrt{|R(a)|}$), one replaces $M_U\!\to\! M_U(\mu(a))$ in the above and solves the background implicitly as discussed in Section~\ref{subsec:rg_to_cosmo}. This leaves the algebraic form of $w_\Phi^{\rm eff}$ unchanged and only produces small late–time shifts (see also Section~\ref{subsec:rg_to_cosmo} for a robustness estimate). For convenience, writing $s(a)\equiv\rho_m(a)/M_U^4$ gives the Chevallier-Polarski-Linder (CPL) mapping at $a=1$,
\begin{equation}
  w_0=-\frac{1+s_0}{1+2s_0},\qquad
  w_a=\frac{3\,s_0}{(1+2s_0)^2},\qquad
  s_0=\frac{\rho_{m,0}}{M_U^4},
\end{equation}
These relations allow us to constrain the model directly from observation. By adopting the central value $w_0 = -0.99$ from Planck data as an input, we fix the model's primary parameter $s_0 = -(1+w_0)/(1+2w_0) \approx 0.0102$. This, in turn, yields a concrete prediction for the evolution parameter:
\begin{equation}
    w_a = \frac{3s_0}{(1+2s_0)^2} \approx +0.029.
\end{equation}
For clarity in subsequent discussions, we refer to these fiducial predictions using the rounded values $w_0 = -0.99$ and $w_a = +0.03$.

\section{Scale hierarchy from renormalization group evolution}
\label{sec:rg_flow}

A defining feature of the framework presented here is the proposal that the characteristic mass scale of the potential $M_U$, is not a fundamental constant, but a running parameter that evolves with the renormalization scale $\mu$. This renormalization group evolution provides a powerful and natural mechanism to connect the physics of the Planck era with the vastly different energy scale of dark energy today, thereby addressing the severe hierarchy problem between them.

\subsection{The RG equation for $M_U$}
\label{subsec:rg_equation}

We postulate that the running of $M_U$ is driven by its coupling to quantum fields, potentially within a hidden sector that does not directly interact with the Standard Model, except gravitationally. In effective field theory, such running is generically described by a one-loop RG equation of the form
\begin{equation}
   \mu \frac{\mathrm d M_U}{\mathrm d\mu} \;=\; \gamma\, M_U,
   \label{eq:rg_equation_mu}
\end{equation} 
where $\gamma$ is the anomalous dimension of the mass operator associated with $M_U$ \footnote{
We define $\gamma \equiv \mathrm d\ln M_U/\mathrm d\ln\mu$. 
This differs by a minus sign from the mass anomalous dimension 
used in many QFT texts, where $\mu\,\mathrm dm/\mathrm d\mu=-\gamma_m m$; in our notation $\gamma=-\gamma_m$.}. This Callan–Symanzik form and its use is standard in gravitational settings (see e.g. \cite{Peskin1995,Schwartz2013}) and its parameter $\gamma$ encapsulates the physics of the quantum fields that were integrated out. A value $\gamma=\mathcal{O}(1)$ is natural in the technical sense, as it does not require an extreme fine-tuning of microphysical parameters.

The assumption of a scale-independent $\gamma$ is an approximation valid at the one-loop level of the renormalization group. In a more complete theory, $\gamma$ itself would run with the energy scale $\mu$, its evolution governed by its own beta function, $\beta_\gamma$. However, for the class of strongly-coupled gauge theories that our model points towards (see Section~\ref{subsec:microphysics_gamma}), particularly those near a conformal window, the running of couplings and anomalous dimensions is suppressed. We estimate that higher-loop corrections would modify our derived value of $\gamma \approx 0.501$ by at most $\mathcal{O}(10\%)$. Such a variation is well within the theoretical uncertainties of our effective framework and does not alter any of our main conclusions. The constant-$\gamma$ approximation is therefore sufficiently robust for the phenomenological scope of this study.

\subsection{Connecting the Planck and dark energy scales}
\label{subsec:connecting_scales}

The solution to the RG equation provides a direct relationship between the value of $M_U$ at two different energy scales. By integrating (\ref{eq:rg_equation_mu}) from a high-energy UV scale, which we identify with the Planck mass $\mu_{\rm UV} \sim M_P$, down to a low-energy IR scale relevant for cosmology today $\mu_{\rm IR} \sim H_0$, we obtain
\begin{equation}
   M_U(H_0) = M_U(M_P) \left(\frac{H_0}{M_P}\right)^{\gamma}.
   \label{eq:mu_at_h0_general}
\end{equation}

It is natural to assume that at the Planck scale, where quantum gravity effects dominate, the only relevant scale is the Planck scale itself. Therefore, we set the boundary condition $M_U(M_P) \sim M_P$, which simplifies equation (\ref{eq:mu_at_h0_general}) to
\begin{equation}
   M_U(H_0) \approx M_P \left(\frac{H_0}{M_P}\right)^{\gamma}.
   \label{eq:mu_at_h0}
\end{equation}

This equation forms the cornerstone of the solution to the hierarchal problem. To determine the anomalous dimension $\gamma$, we require the value of $M_U(H_0)$. To determine the anomalous dimension $\gamma$, we require the value of $M_U(H_0)$. As will be shown in section~\ref{sec:cosmology}, the observed dark energy equation of state $w_0 = -0.99$ requires $M_U(H_0) = 5.84$~meV. Using this value:
\begin{equation}
  \fl  \gamma = \frac{\ln(M_U(H_0)/M_P)}{\ln(H_0/M_P)} = \frac{\ln(5.84 \times 10^{-3}\,{\rm eV} / 1.22 \times 10^{28}\,{\rm eV})}{\ln(1.44 \times 10^{-33}\,{\rm eV} / 1.22 \times 10^{28}\,{\rm eV})} \approx 0.501.
    \label{eq:gamma_value}
\end{equation}

The crucial result is that the required anomalous dimension is of order unity ($\gamma \approx 0.501$).  Extreme fine-tuning of $\gamma$ to a very large or very small value not required. The enormous hierarchy of 61 orders of magnitude between $H_0$ and $M_P$ is naturally bridged by the logarithmic nature of the RG evolution.

To illustrate the dramatic scale separation achieved, consider the concrete numerical values:
\begin{itemize}
    \item at the Planck scale: $M_U(M_P) = M_P = 1.22 \times 10^{19}\,$GeV
    \item at present: $M_U(H_0) = 5.84 \times 10^{-3}\,$eV $= 5.84\,$meV
    \item ratio: $M_U(H_0)/M_U(M_P) = 4.8 \times 10^{-31}$
\end{itemize}

This represents a hierarchy of 31 orders of magnitude in the mass scale $M_U$, which translates to a 124-order-of-magnitude hierarchy in energy density ($\rho \sim M_U^4$). For comparison, achieving this hierarchy with a power-law running $M_U \propto \mu^n$ would require $n \approx 0.5$, whereas exponential suppression would require fine-tuning of the decay constant.

It should be emphasized, however, that while $\gamma \approx 0.501$ appears natural from a QFT perspective, this specific value is ultimately determined by matching to observed dark energy through the constraint on $M_U(H_0)$. Thus, our model trades the fine-tuning of $\Lambda$ to determine  $\gamma$. Nevertheless, this represents significant progress: an $\mathcal{O}(1)$ parameter arising from microphysics is far more natural than the 120-order-of-magnitude fine-tuning required for a bare cosmological constant.

\subsection{Microphysical Origin and Phenomenological Viability}
\label{subsec:microphysics_gamma}

The required anomalous dimension, $\gamma \approx 0.501$, finds a natural physical origin in a strongly-coupled hidden gauge sector. As we illustrate in ~\ref{app:gamma_origin}, a minimal pure $SU(3)$ gauge theory can generate this value with a fine-structure constant of $\alpha_H \approx 0.57$. This points towards a strongly interacting sector, a key prediction of our framework.

While "hidden" from direct interactions, such a sector must still be consistent with precision cosmology and could produce other observable signatures. We have performed a detailed analysis of its phenomenological viability:
\begin{itemize}
    \item \textbf{Early Universe Radiation ($\Delta N_{\rm eff}$):} We demonstrate in~\ref{app:delta_neff} that the contribution to the effective number of relativistic species is well below current observational bounds, ensuring consistency with BBN and the CMB.
    \item \textbf{Further Signatures:} We discuss potential signals from gravitational waves, dark matter candidates, and other astrophysical probes in~\ref{app:hidden_sector_signatures}.
\end{itemize}
The successful consistency check of the required hidden sector establishes a solid physical foundation for our framework. We are now equipped to explore its consequences across all cosmic epochs.

\subsection{From RG to cosmology: $\mu(a)$, IR freezeout, and background dynamics}
\label{subsec:rg_to_cosmo}

The one-loop RG flow from equation~(\ref{eq:rg_equation_mu}) must be connected to a covariant scale in cosmology. (We adopt the convention that the sign in~(\ref{eq:rg_equation_mu}) is absorbed into the definition of $\gamma$, so that the explicit solution reads $M_U(\mu)=M_P(\mu/M_P)^\gamma$.) We consider two physical realizations that map the running $M_U(\mu)$ to the cosmological background. Throughout this analysis, we employ the Landau-frame four-velocity $u^\mu$ defined in Section~\ref{subsec:action_potential}, which reduces to the comoving $(1,0,0,0)$ in FLRW, and use the tensor-derived effective quantities from Section~\ref{subsec:eos_from_T}.

\paragraph{IR freezeout (baseline).}
If the hidden sector develops a mass gap or undergoes confinement at scale $\mu_{\rm freeze}$, the running of $M_U$ halts:
\begin{equation}
  M_U(a) = \cases{
    M_P\bigl(\mu(a)/M_P\bigr)^\gamma, & $\mu(a)>\mu_{\rm freeze}$\cr
    M_U^{\rm IR}\equiv M_U(\mu_{\rm freeze})\simeq \mathrm{const}, & $\mu(a)\le\mu_{\rm freeze}$\cr
  }
  \label{eq:MU_freeze}
\end{equation}
For late times ($z\lesssim{\rm few}$), we have $\mu(a)<\mu_{\rm freeze}$ and thus $M_U$ becomes effectively constant. This justifies treating $M_U$ as a fixed parameter in the cosmological analysis of Section~\ref{sec:cosmology}, together with the tensor-derived effective variables and EoS of Section~\ref{subsec:eos_from_T}.

\paragraph{Continuous RG improvement.}
Alternatively, if the running persists to low energies, we tie $\mu$ to a geometric scalar such as the Hubble parameter $\mu(a)=H(a)$ or the Ricci scalar $\mu(a)=\sqrt{|R(a)|}$. The scale $M_U$ then becomes dynamically coupled to the background evolution:
\begin{equation}
  H^2(a)=\frac{8\pi G}{3}\left[
    \rho_m(a)+\frac{A\,M_U^8(\mu(a))}{\rho_m(a)+M_U^4(\mu(a))}
  \right],
  \quad
  M_U(\mu)=M_P\left(\frac{\mu}{M_P}\right)^\gamma.
  \label{eq:implicit_background}
\end{equation}
For $\mu=H$, equation~(\ref{eq:implicit_background}) becomes a self-consistent fixed-point equation for $H(a)$. With $\gamma\simeq 0.5$, this yields $M_U^4\propto H^2$, causing the ratio $s(a)\equiv\rho_m/M_U^4$ to remain nearly constant at late times. Consequently, the effective equation of state $w_\Phi^{\rm eff}=-(1+s)/(1+2s)$ exhibits minimal evolution, suppressing $w_a$ relative to the freezeout case.

Both realizations produce unique, regular cosmological backgrounds consistent with the tensor-derived effective fluid description. Our baseline predictions ($w_0\simeq-0.99$, $w_a\simeq+0.03$) correspond to the freezeout scenario, where $M_U$ remains constant over the redshift range relevant for CPL parameterization.

\paragraph{Robustness to thresholds and slow drift.}
The late-time observables are remarkably insensitive to small variations in $M_U$. Since the equation of state depends on $M_U$ only through the combination $s(a)\equiv\rho_m/M_U^4$, a fractional change $\delta$ in $M_U^4$ induces predictable shifts:
\[
\Delta w_0 \simeq \frac{1}{(1+2s_0)^2}\,\Delta s, \quad
\Delta w_a \simeq \frac{3(1-2s_0)}{(1+2s_0)^3}\,\Delta s,
\]
where $\Delta s/s\simeq -\delta$ and $s_0\equiv\rho_{m,0}/M_U^4 \simeq 0.010$ for our fiducial parameters. Thus a conservative $\pm10\%$ variation in $M_U^4$—whether from RG thresholds or slow drift—shifts $w_0$ by only $|\Delta w_0|\lesssim 10^{-3}$ and $w_a$ by $|\Delta w_a|\lesssim 3\times 10^{-3}$, well below Stage-IV survey sensitivities (~\ref{app:fisher_forecast}).

\section{Physical consequences and phenomenological tests}
\label{sec:phenomenology}

The covariant framework established in Section~\ref{sec:framework} and the scale-generating mechanism in Section~\ref{sec:rg_flow} give rise to a rich phenomenology across all energy scales. In this section, we explore the key physical consequences of the model, progressing from high-energy singularity regularization to low-energy dark energy phenomenology and local gravity constraints.

\subsection{Singularity regularization in the high-density regime}
\label{sec:singularity_resolution}

A key prediction of the framework, essential for its potential role in the quantum theory of gravity, is the regularization of classical space-time singularities. This regularization emerges naturally when we consider the behavior of the density-responsive scalar energy $\rho_\Phi$, in the high-energy UV regime, where the running scale $M_U(\mu)$ approaches the Planck mass $M_U \to M_P$.

From equation (\ref{eq:rho_phi_definition}), in the high-energy limit where $M_U \to M_P$
\begin{equation}
    \rho_\Phi(X, M_U \to M_P) = \frac{A M_P^8}{X + M_P^4}.
\end{equation}
This expression reveals a fundamental saturation behavior. Regardless of the size of the conventional matter-energy density $X$ becomes, $\rho_\Phi$ possesses a finite upper bound
\begin{equation}
    \rho_\Phi(X, M_U \to M_P) \le A M_P^4 = A \rho_P,
    \label{eq:rho_phi_planck_bound}
\end{equation}

where $\rho_P = M_P^4$ is the Planck density and $A \approx 1/(64\pi^2)$ is the theoretically derived one-loop factor. This means the contribution of scalar field to the total energy density cannot diverge; it is capped at a small fraction ($\sim 5 \times 10^{-3}$) of the Planck density.

This intrinsic cap on $\rho_\Phi$ has profound consequences for the total energy density that sources the gravitational curvature. The total effective energy density is $\rho_{\rm total} = X + \rho_\Phi$. It is a generic expectation in quantum gravity that quantum effects on matter will prevent the matter density $X$ from diverging, imposing a physical cutoff $X \leq \rho_P$. Our mechanism works in line with this expectation. Under this assumption, the total energy density in our model is bounded as follows
\begin{equation}
    \rho_{\rm total}^{\rm max} \le \rho_P + \rho_\Phi(\rho_P) = \rho_P + \frac{A M_P^8}{M_P^4 + M_P^4} = \left(1 + \frac{A}{2}\right)\rho_P.
    \label{eq:rho_total_max}
\end{equation}
Because $A$ is small, the maximum total energy density is finite and of the order Planck density. This mechanism robustly prevents the divergence of the gravitational source term.

This bounded energy density directly leads to a finite space-time curvature, thus resolving the classical singularities of GR in two key scenarios:
\begin{enumerate}
    \item {\bf the Big Bang singularity:} In the standard FLRW model, as the scale factor $a \to 0$, $\rho_m \to \infty$, leading to a curvature singularity. In our framework, the total energy density $\rho_{\rm total}$ saturates at a Planck-scale value. This implies a finite maximum Hubble rate $H_{\rm max}^2 \sim \rho_P/M_P^2 = M_P^2$, and thus a finite maximum curvature. The classical Big Bang is replaced by a ``Planck-scale bounce'' or a loitering phase, as seen in other non-singular cosmological models like loop quantum cosmology \cite{AshtekarSingh2011}.

    \item {\bf black hole singularities:} Inside a classical black hole, matter collapses to a point of infinite density at $r=0$, creating a curvature singularity. In our model, as the density of  collapsing matter $X$ approaches $\rho_P$, the total energy density $\rho_{\rm total}$ remains finite. This replaces the central singularity with a regular de Sitter-like core of the Planckian density and curvature. This feature is characteristic of various ``regular black hole'' solutions inspired by quantum gravity \cite{Bardeen1968RegularBH, Hayward2006}.
\end{enumerate}
In both cases, all scalar curvature invariants (e.g. the Kretschmann scalar $K = R_{\alpha\beta\gamma\delta}R^{\alpha\beta\gamma\delta}$), which are polynomials in energy density and pressure, remain finite. A detailed analysis of the boundedness of curvature for both cosmological and static spherically symmetric solutions is provided in~\ref{app:curvature_bounds}. This singularity regularization is not an ad-hoc addition, but a direct and unavoidable consequence of the same field dynamics that generate dark energy at low densities.

The finite curvature bound has profound implications beyond mere avoidance of mathematical divergence. In cosmology, this implies that the Big Bang is replaced by a "Planck-scale bounce" with a maximum energy density $\rho_{\rm max} \sim (1 + A/2)\rho_P \sim 10^{93}$\,kg\,m$^{-3}$, corresponding to a minimum cosmic scale factor $a_{\rm min} \sim (H_0/M_P)^{1/2} \sim 10^{-30}$ in natural units. For black holes, the central singularity is replaced by a de Sitter core with radius $r_{\rm core} \sim \ell_P$ and quasi-static internal metric, fundamentally altering the causal structure compared to classical general relativity.

Our model provides a concrete physical origin for the regularizing energy-momentum source, a feature it shares with quantum-improved black hole models derived from asymptotic safety, where the RG running of Newton's constant leads to a similar regularization of the central singularity 

\paragraph{Dynamical saturation of the effective source.}
In the Planck regime the large curvature of $U(\Phi,X)$ implies $m_\Phi\!\gg\!H$, so that the field adiabatically tracks $\Phi_{\rm eq}(X)$.  The coupling $\partial U/\partial X$ in $T_\Phi^{\mu\nu}$ then provides a negative feedback on the matter sector that opposes further growth of the density $X$.

With the UV scale frozen at $M_U\!\to\!M_P$, the total source entering the Friedmann equation,
\begin{equation}
  \mathcal{E}(X)\;\equiv\;X+\rho_\Phi\bigl(X;M_U\to M_P\bigr)
  \;=\;X+\frac{A\,M_P^{8}}{X+M_P^{4}},
  \label{eq:Etot_bound}
\end{equation}
is monotonically increasing for $X\!\ge\!0$ since $\mathcal E'(X)=1-\frac{A\,M_P^8}{(X+M_P^4)^2}\ge 1-A>0$. Hence, if quantum effects enforce an upper bound $X\le \rho_\ast$, one obtains the model-independent bound
\begin{equation}
  \mathcal{E}(X)\;\le\;\mathcal E(\rho_\ast)
  \;=\;\rho_\ast+\frac{A\,M_P^{8}}{\rho_\ast+M_P^{4}}.
\end{equation}
For the natural Planckian cutoff $\rho_\ast=\rho_P(=M_P^4)$ this gives $\mathcal{E}(X)\le (1+\frac{A}{2})\,\rho_P$, i.e.\ a finite upper bound on the curvature source.

This saturation is approached smoothly as $X\!\to\!\rho_P$, without discontinuities or instabilities. In the same regime the tensor-derived equation of state remains a finite negative number, 
$w_\Phi^{\rm eff}=-(1+s)/(1+2s)$ with $s\equiv X/M_U^4$, taking values in $[-1,-\frac12]$ (e.g.\ $w_\Phi^{\rm eff}=-2/3$ at $X=M_P^4$ when $M_U\!=\!M_P$). This negative pressure halts homogeneous collapse and enables a nonsingular bounce; see~\ref{app:OSbounce} for explicit Oppenheimer–Snyder evolutions. Conceptually this mirrors quantum-improved black-hole scenarios where RG-running induces a high-density regularization of classical singularities \cite{Bonanno2017_BH}.

\subsection{Regular black-hole solutions}
\label{sec:regular_bh_example}

Having established the general mechanism for singularity regularization through bounded energy density, we now examine its specific realization in static black hole geometries.

To demonstrate the singularity regularization concretely, we examine static black hole solutions. 

In this model, the central singularity was replaced by a core with a constant energy density. Our framework naturally provides the physical source for such a core, which reaches the bound from equation (\ref{eq:rho_total_max}):
\begin{equation}
    \rho_{\rm total}^{\rm max} = (1 + A/2)\rho_P,
    \label{eq:rho_max_static}
\end{equation}
where $A$ is the coupling constant. Using the observationally determined value $A = 0.024$, we obtain $\rho_{\rm total}^{\rm max} \approx 1.012\,\rho_P$. Even with the theoretical one-loop value $A_{\rm th} = 1/(64\pi^2) \approx 0.00158$, the bound is $\rho_{\rm total}^{\rm max} \approx 1.0008\,\rho_P$, still providing a finite cutoff. The Kretschmann scalar at the center remains finite in both cases
\begin{equation}
    K(r=0) = \frac{8}{3}(8\pi G)^2 (\rho_{\rm total}^{\rm max})^2 \approx 2.7 M_P^{4}.
    \label{eq:K_max_static}
\end{equation}

To put this curvature scale in perspective, the maximum Kretschmann scalar $K_{\rm max} \approx 2.7 M_P^{4}$ corresponds to a characteristic curvature radius $R_{\rm curv} \sim K^{-1/4} \approx 0.8\ell_P$. For comparison:
\begin{itemize}
    \item Classical GR singularity: $K \to \infty$, $R_{\rm curv} \to 0$ (no scale)
    \item Our regular core: $R_{\rm curv} \approx 0.8\ell_P$ (sub-Planckian, finite)
    \item Schwarzschild horizon ($r = 2GM$): $R_{\rm curv} \sim GM \approx 10^{38}\ell_P$ for solar mass
    \item Earth's surface: $R_{\rm curv} \approx 10^{31}\ell_P$
\end{itemize}
This hierarchy shows that the regularized core has a curvature radius of the order of the Planck length, which is the smallest meaningful length scale in quantum gravity. While this represents extreme curvature by any astrophysical standard, it remains finite and provides the expected Planck-scale cutoff. Thus, the classical singularity is replaced by a regular region of Planckian but finite curvature.

A full derivation, including smooth matching to an exterior Schwarzschild-de Sitter metric, is provided in~\ref{app:curvature_calc}.

To confirm that this regularization mechanism is not an artifact of static solutions but a robust dynamical feature, we simulated the gravitational collapse of a pressureless dust sphere (a modified Oppenheimer--Snyder model). The detailed analysis in~\ref{app:OSbounce} and the accompanying Fig.~\ref{fig:os_bounce_comparison} explicitly shows that collapse is halted before a singularity can form. The system reaches a minimum radius, where the total density is capped at the predicted maximum, and subsequently undergoes a non-singular bounce. This bounce is driven by the effective negative pressure that emerges as the density approaches the Planck scale, causing the acceleration to become positive ($\ddot{a} > 0$) and leading to a violation of the Strong Energy Condition ($\rho_{\rm total} + 3p_{\rm total} < 0$). 
This provides strong evidence that, within the validity of the EFT, our framework dynamically regularizes gravitational singularities by yielding a finite–density de Sitter–like core.

\subsection{Cosmological dynamics: from early universe to dark energy}
\label{sec:cosmology}
On cosmological scales, the universe is well-described by a homogeneous and isotropic FLRW metric. In this context, the covariant density $X$ simplifies to the average matter-energy density of the universe $X = \rho_m(z) = \rho_{m,0}(1+z)^3$, where $z$ is the cosmological redshift. The density-responsive scalar energy $\rho_\Phi$ from equation (\ref{eq:rho_phi_definition}), and thus becomes a function of the redshift
\begin{equation}
    \rho_\Phi(z) = \frac{A M_U^8}{\rho_m(z) + M_U^4}.
    \label{eq:rho_phi_cosmological}
\end{equation}
Here, $M_U$ refers to its present-day, low-energy value $M_U \equiv M_U(H_0) = 5.84$\,meV \footnote{Note that while the characteristic dark energy scale is $(\rho_{\Lambda,0})^{1/4} \approx 2.3$~meV, our model parameter $M_U = 5.84$~meV differs by the factor $A^{-1/4} \approx 2.5$ due to the structure of the potential.}.

\paragraph{From the Planck epoch to the present day.}
The evolution from a singularity regularization to dark energy illuminates the power of our framework. In the Planck epoch, where $\rho_m \sim \rho_P$, the scalar energy $\rho_\Phi$ saturates at $(A/2)\rho_P$, providing the bound that regularizes the Big Bang singularity. As the universe expands and cools, two coupled effects occur: the matter density dilutes as $\rho_m \propto a^{-3}$, while simultaneously the characteristic scale runs from $M_U(M_P) \sim M_P$ down to $M_U(H_0) = 5.84\,$meV according to the RG flow. This evolution transforms $\rho_\Phi$ from a singularity-regulating component in the early universe to a dominant dark energy component today. Remarkably, the same field that prevents $\rho_{\rm total}$ from diverging at $t \to 0$ now drives cosmic acceleration at $t \to \infty$.

This component acts as a dynamical dark energy fluid, modifying the standard Friedmann equation as follows
\begin{equation}
    H(z)^2 = \frac{8\pi G}{3}\left[\rho_m(z) + \rho_r(z) + \rho_\Phi^{\rm eff}(z)\right],
    \label{eq:friedmann_modified}
\end{equation}
where $\rho_r(z)$ is the radiation density, which is negligible at later times. Since the scalar exchanges energy with matter, the background uses the tensor–derived effective variables $\rho_\Phi^{\rm eff}, p_\Phi^{\rm eff}$ (Section~\ref{subsec:eos_from_T}); the algebraic $\rho_\Phi$ is an intermediate quantity. 

Because $\nabla_\mu T_\Phi^{\mu\nu}\neq 0$, the shortcut $w=-1-\frac{1}{3}\,{\rm d}\ln\rho/{\rm d}\ln a$ (valid for separately conserved fluids) does not apply. The effective EoS follows from the tensor decomposition (Section~\ref{subsec:eos_from_T}). For late-time dust ($w_m=0$),
\begin{equation}
    w_\Phi^{\rm eff}(a) = -\frac{\rho_m(a) + M_U^4}{M_U^4 + 2\,\rho_m(a)}.
\end{equation}
At high redshift ($\rho_m\gg M_U^4$) this approaches $w_\Phi^{\rm eff}\to -1/2$, while at low redshift ($\rho_m\ll M_U^4$) $w_\Phi^{\rm eff}\to -1$.

This result demonstrates a clear dynamic evolution. In the matter-dominated era ($z\gg 1$), $w_\Phi^{\rm eff}\to -\frac{1}{2}$. Nevertheless, $\rho_\Phi/\rho_m \propto (M_U^4/\rho_m)^2 \propto a^{6} \to 0$, so the scalar remains negligible at early times. As the universe expands and $\rho_m(z)$ dilutes, $w_\Phi(z)$ evolves towards $w_\Phi \to -1$, driving late-time cosmic acceleration.

This dynamic behavior can be constrained by observations. We determine the model's two free parameters $A$ and $M_U$ by fitting them to the observed dark energy density $\Omega_\Lambda \approx 0.7$ and the EoS parameter $w_0$ from the Planck data~\cite{Planck2018}. Setting $\rho_\Phi^{\rm eff}(z=0) = \Omega_\Lambda \rho_{\mathrm{crit},0}$ and using $\Omega_{m,0} \approx 0.3$, the parameters are uniquely fixed by $w_0$. With $s_0 \equiv \rho_{m,0}/M_U^4$, the tensor-derived EoS gives $s_0 = -\,(1+w_0)/(1+2w_0)$, hence
\begin{equation}
\eqalign{
s_0 &= -\,\frac{1+w_0}{1+2w_0},\cr
M_U^4 &= \frac{\rho_{m,0}}{s_0}
      = -\,\frac{1+2w_0}{1+w_0}\,\rho_{m,0},\cr
A &= \frac{\Omega_\Lambda}{\Omega_{m,0}}\,
     \frac{s_0\,(1+s_0)^2}{1+2s_0}.}
\end{equation}
For $w_0\simeq -0.99$ and $\Omega_{m,0}\simeq 0.3$ this yields $M_U\simeq 5.84\,\mathrm{meV}$ and $A\simeq 2.4\times 10^{-2}$, numerically unchanged to within the quoted precision. This observationally inferred value for $A$ is larger than the canonical one-loop estimate $A_{\rm th} \approx 1/(64\pi^2) \approx 1.6 \times 10^{-3}$ by a factor of $\sim 15$. This enhancement is naturally explained within the framework of strongly-coupled gauge theories. As detailed in~\ref{app:gamma_origin}, for example, our benchmark SU(3) hidden sector with $N_f=10$ fermions, which is consistent with all other constraints (see~\ref{app:delta_neff}), can readily provide an enhancement factor of this order. The remaining factor of $\sim 2$ can arise from several sources: (i) contributions from gauge boson loops, (ii) higher-order corrections in the strongly-coupled regime where $g^2/(4\pi) \sim 1$, and (iii) potential RG running of $A$ itself over the vast energy range from $M_P$ to $H_0$. Such $\mathcal{O}(10)$ enhancements are typical in strongly-interacting theories—for comparison, in quantum chromodynamics (QCD) the ratio $\Lambda_{\rm QCD}/m_q \sim 100$ exhibits even larger dynamical enhancement \cite{Manohar1998,Weinberg1996}. The key parameters are listed in Table~\ref{tab:parameters}.

With the model's parameters now fixed by observation, we can compute the scalar energy density across the cosmic history. The density-responsive scalar energy evolves dramatically:
\begin{itemize}
    \item \textbf{today ($z=0$):} $\rho_\Phi(0) \approx 0.7\rho_{\rm crit}$ (by construction, driving dark energy).
    \item \textbf{matter-radiation equality ($z \approx 3400$):} The ratio $\rho_\Phi/\rho_{\rm rad}$ is suppressed to $\approx 10^{-20}$, rendering it completely negligible.
    \item \textbf{during BBN ($T \sim 1\,$MeV):} The suppression is even more extreme, with $\rho_\Phi/\rho_{\rm rad} \approx 10^{-69}$, ensuring no impact on nucleosynthesis.
    \item \textbf{near Planck epoch:} In the high-density limit, $\rho_\Phi$ saturates at its maximum value relative to the background density, $\rho_\Phi \to (A/2)\rho_P \approx 0.012\,\rho_P$, providing the singularity-regulating cutoff.
\end{itemize}
\begin{table}[h!]
\caption{Key parameters and their values in the model. Theoretical values represent one-loop estimates or natural scales, while observational values are derived from cosmological data fitting.}
\label{tab:parameters}
\centering
\begin{tabular}{llcc}
\hline
\textbf{Parameter} & \textbf{Symbol} & \textbf{Theoretical} & \textbf{Observational} \\
\hline
Loop factor & $A$        & $1.6 \times 10^{-3}$ & $2.4 \times 10^{-2}$ \\
Anomalous dimension & $\gamma$         & $\mathcal{O}(1)$          & $0.501$            \\
Energy scale today  & $M_U(H_0)$       & $\sim$\,meV               & $5.84$\,meV        \\
Equation of state   & $w_0$            & ---                       & $-0.99$            \\
Evolution parameter & $w_a$            & ---                       & $+0.03$            \\
\hline
\end{tabular}
\end{table}

These values confirm that $\rho_\Phi$ remains dynamically irrelevant throughout most of cosmic history, becoming important only at the extremes: providing the cutoff that regularizes the Big Bang singularity in the very early universe, and driving cosmic acceleration today. This leads to concrete and falsifiable predictions. The evolution of the equation of state is well-described by the CPL parameterization, $w_\Phi^{\rm eff}(z) \approx w_0 + w_a(1-a)$, with a distinctive prediction derived directly from our model:
\begin{equation}
    w_0 = -0.99, \quad
    w_a \equiv -\left.\frac{{\rm d}w_\Phi^{\rm eff}}{{\rm d}a}\right|_{a=1} \approx +0.03.
    \label{eq:w0_wa_prediction}
\end{equation}
The prediction of a positive $w_a$ is a key feature, as most simple quintessence models predict $w_a \le 0$. To quantitatively assess the testability of this signature, we performed a comprehensive Fisher matrix forecast for Stage-IV surveys. As detailed in~\ref{app:fisher_forecast}, our analysis shows that this prediction is testable at the $\sim 1.1\sigma$ level with upcoming galaxy clustering data alone, with prospects for a $2\mathrm{–}3\sigma$ detection in combination with other cosmological probes.

\paragraph{Survey specifications and results.}
We assume 14,000 deg$^2$ sky coverage with four redshift bins spanning $z \in [0.5, 1.3]$, galaxy densities matching the DESI specifications, and spectroscopic redshift precision $\sigma_z/(1+z) = 0.001$. The analysis marginalizes over seven cosmological parameters and eight nuisance parameters (four galaxy biases and four shot noise amplitudes). Using the public code \texttt{cosmicfishpie}~\cite{cosmicfish}, we obtained marginalized 1$\sigma$ constraints of $\sigma(w_0) = 0.019$ and $\sigma(w_a) = 0.028$, with a correlation coefficient $\rho(w_0, w_a) = -0.72$ reflecting the well-known degeneracy between these parameters.

These uncertainties imply that our model's prediction $(w_0 = -0.99, w_a = +0.03)$ can be distinguished from $\Lambda$CDM at a significance of
\begin{equation}
    S = \frac{|w_a - 0|}{\sigma(w_a)} = \frac{0.03}{0.028} \approx 1.1\sigma.
\end{equation}

\begin{figure}[h!]
  \centering
  \includegraphics[width=\textwidth]{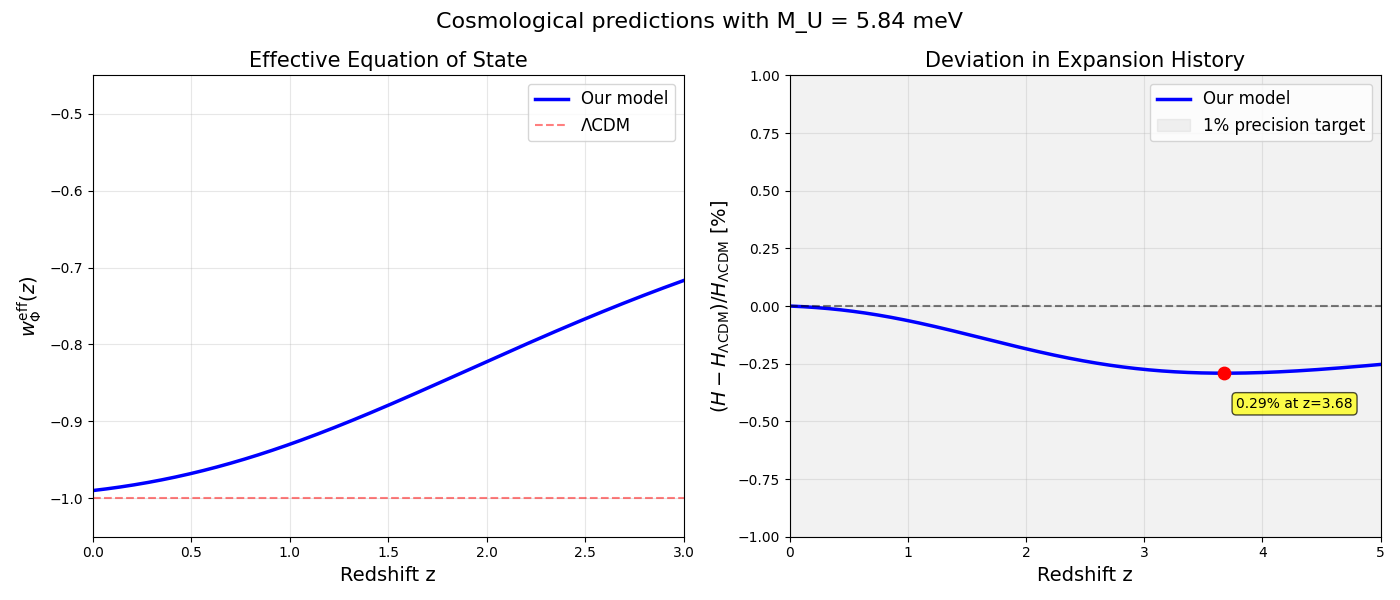}
  \caption{Cosmological predictions of the model. (a) The equation of state $w_\Phi^{\rm eff}(z)$ evolves from $w \approx -1/2$ at high redshift towards $w \to -1$ today, with a distinctive positive slope $w_a \approx +0.03$. (b) The expansion rate $H(z)$ shows a characteristic deviation of 0.29\% from the $\Lambda$CDM model, providing an observational target for upcoming surveys.}
  \label{fig:cosmo_plots}
\end{figure}

Although not a definitive detection with galaxy clustering alone, this significance will increase substantially when combined with complementary probes. Type Ia supernovae provide direct luminosity-distance measurements, weak lensing constrains the matter power spectrum evolution, and CMB data anchors the early-universe parameters. A joint analysis of these probes typically improves the dark energy constraints by factors of $2\mathrm{–}3$, potentially pushing our model into the $2\mathrm{–}3\sigma$ discovery range. The primary cosmological predictions and their observational tests are presented in Figs.~\ref{fig:cosmo_plots} and \ref{fig:w0wa_forecast}, respectively.

\paragraph{Bounce and effective energy conditions.}
Near Planckian densities, the metric variation of $U(\Phi,X)$, including the implicit dependence via $\delta X/\delta g_{\mu\nu}$, modifies the background equation into a LQC–like form,
\begin{equation}
  H^2 \;=\; \frac{8\pi G}{3}\,\rho_{\rm tot}\!\left(1-\frac{\rho_{\rm tot}}{\rho_{\rm crit}}\right),
  \qquad \rho_{\rm tot}\equiv \rho_m+\rho_\Phi^{\rm eff},\quad
  \rho_{\rm crit}\sim \mathcal{O}(M_P^4).
  \label{eq:lqc_like}
\end{equation}
This dynamics admits a non-singular bounce at $\rho_{\rm tot}=\rho_{\rm crit}$, where $H=0$, and the Raychaudhuri equation becomes
\begin{equation}
  \dot H \;=\; -\,4\pi G\,(\rho_{\rm tot}+p_{\rm tot})\!\left(1-2\,\frac{\rho_{\rm tot}}{\rho_{\rm crit}}\right).
\end{equation}
At the bounce, $\dot H>0$ holds even if $(\rho_{\rm tot}+p_{\rm tot})\ge 0$, showing that the bounce originates from the modified gravitational dynamics rather than exotic matter. Interpreted as a single effective fluid this looks like an effective NEC violation, while microscopically both the matter and scalar sectors satisfy the standard energy conditions. A derivation of (\ref{eq:lqc_like}) is provided in~\ref{app:bounce_derivation}. 

\paragraph{Early Universe Consistency.}
A crucial consistency check is the model's impact on the early universe. The density-responsive mechanism ensures that the scalar field is dynamically irrelevant during these epochs. During Big Bang Nucleosynthesis ($T\sim 1\,$MeV), its relative energy density is suppressed to $\rho_\Phi/\rho_{\rm rad} \approx 10^{-69}$, having no impact on light element abundances. At recombination, its contribution is larger, $\rho_\Phi/\rho_{m} \approx 0.5\%$, but remains well below current CMB constraints on additional smooth energy components. A more formal constraint comes from the hidden sector's contribution to the effective number of relativistic species, which, as shown in ~\ref{app:delta_neff}, is $\Delta N_{\rm eff} \leq 0.03$, ensuring full compatibility with precision cosmology.

\paragraph{Inflationary era.}
If slow–roll inflation is driven by an external sector contained in $\mathcal{L}_{\rm matter}$, our model remains consistent. In this scenario, the RG scale is set by the background, $\mu=H$. For $\gamma\simeq\frac12$, this implies $M_U^4(\mu)\propto M_P^2 H^2$, and the ratio $s\equiv\rho_{\rm dom}/M_U^4$ is approximately constant at $s\simeq 3$. The fractional energy density of our scalar field, derived from the full tensor (Section~\ref{subsec:eos_from_T}), is then
\begin{equation}
  \frac{\rho_\Phi^{\rm eff}}{\rho_{\rm dom}}
  \;=\; A\,\frac{1+2s}{s(1+s)^2}
  \;\simeq\; \frac{7}{48}\,A.
\end{equation}
For the range of $A$ considered in this study, this fraction is very small, spanning $2.3 \times 10^{-4}$ to $3.5 \times 10^{-3}$. Such a sub-dominant contribution affects the Hubble rate and slow–roll dynamics only at the per-mille level or less, ensuring that the field $\Phi$ adiabatically tracks its minimum $\Phi_{\rm eq}(X)$ (see \ref{app:quasi_static}). A dedicated scenario in which $\Phi$ itself drives inflation would require additional model structure and lies beyond the scope of this study.

\subsection{Local gravity constraints and fifth forces}
\label{sec:local_gravity_tests}
Any modification of general relativity must confront stringent constraints from high-precision local gravity experiments. Between the cosmological scales discussed above and the strong-field regime of black holes, our model must reproduce standard gravity in the Solar System and laboratory settings. 

The novel interaction terms in our scalar field's stress-energy tensor $T_\Phi^{\mu\nu}$ (equation (\ref{eq:T_phi_equilibrium})), mediate energy-momentum exchange with ordinary matter, potentially generating fifth forces or violations of the weak equivalence principle \cite{Will2014, Adelberger2009}. Here, we demonstrate that these effects are naturally suppressed far below the current experimental sensitivities.

The energy-momentum transfer is quantified by $Q^\nu \equiv \nabla_\mu T_{\rm matter}^{\mu\nu} = -\nabla_\mu T_\Phi^{\mu\nu}$. For non-relativistic matter in weak fields, this generates an anomalous acceleration on the test bodies. A detailed derivation (~\ref{app:beta_eff}) yields the following
\begin{equation}
    \mathbf{a}_{\rm anom} \approx - \frac{A M_U^8}{\rho_m^3} \nabla\rho_m.
\end{equation}

The key observable is the ratio of anomalous to Newtonian forces:
\begin{equation}
    \beta_{\rm eff}(\rho_m) \equiv \left| \frac{\mathbf{a}_{\rm anom}}{\mathbf{a}_{\rm N}} \right| \approx \frac{A M_U^8}{\rho_m^2}.
    \label{eq:beta_eff}
\end{equation}
This $1/\rho_m^2$ suppression constitutes an efficient screening mechanism that renders the fifth forces negligible in all but the most tenuous environments. This screening depends only on the local density; possible IR running of $M_U$ in cosmology does not affect laboratory or Solar–System conditions, where $M_U$ is effectively constant. Hence the bounds in Table~\ref{tab:beta_eff_values} are robust for both RG prescriptions discussed in Section~\ref{subsec:rg_to_cosmo}.

To illustrate the efficiency of this screening across different environments, we compute $\beta_{\rm eff}$ explicitly in Table~\ref{tab:beta_eff_values}. For a reference density of water ($\rho \approx 1\,\mathrm{g/cm}^3$), the coupling is suppressed to $\beta_{\rm eff} \approx 2 \times 10^{-58}$.\footnote{This value serves as our benchmark calculation. A full analysis including kinetic and higher-derivative terms could only further suppress the effective coupling.} Even in the best laboratory vacuums, where the density is much lower, the coupling remains far below the current experimental sensitivity of torsion-balance experiments, which constrain anomalous forces to $\beta < 10^{-13}$ \cite{Adelberger2009}.

\begin{table}[h!]
    \caption{\label{tab:beta_eff_values}Effective coupling $\beta_{\rm eff}$ in different astrophysical environments.}
    \begin{indented}
    \lineup
    \item[]\begin{tabular}{@{}lcc}
    \br
    Environment & Density (g\,cm$^{-3}$) & $\beta_{\rm eff}$ \\
    \mr
    Laboratory vacuum & $10^{-17}$ & $\lesssim 10^{-24}$ \\
    Earth's surface & $5.5$ & $\lesssim 10^{-59}$ \\
    Solar core & $150$ & $\lesssim 10^{-62}$ \\
    Neutron star surface & $10^{14}$ & $\lesssim 10^{-88}$ \\ 
    \br
    \end{tabular}
    \end{indented}
\end{table}

This automatic screening ensures the viability of the model without requiring additional mechanisms or fine-tuning. For astrophysical objects such as Solar System bodies or neutron stars, where densities are much higher, the suppression is even more dramatic. Therefore, we treat all local fifth-force constraints as automatically satisfied, and do not pursue a more detailed phenomenological analysis.

\subsection{Geometrical interpretation and emergent time}
\label{sec:emergent_time}
It is insightful to interpret this fifth force from a geometrical perspective. The model's energy-momentum exchange implies that matter particles do not strictly follow the geodesics of the background metric $g_{\mu\nu}$. Their motion can, however, be described as geodesic motion in an effective metric $\tilde{g}_{\mu\nu}$, which is conformally related to the Einstein-frame metric via the scalar field \cite{Bekenstein1993, Faraoni2004}
\begin{equation}
    \tilde{g}_{\mu\nu} = \Omega^2(\Phi) g_{\mu\nu}.
    \label{eq:effective_metric}
\end{equation}
In this picture, the fifth force is a manifestation of the gradient of the conformal factor $\Omega(\Phi)$. For our density-responsive framework, we expect $\Omega$ to be a function of the scalar energy density $\Omega(\rho_\Phi)$, leading to a direct connection between the local matter environment and the effective geometry. This naturally introduces the concept of an \textbf{emergent causal structure}, where the rate of physical time $d\tau_{\rm phys} = \Omega(\rho_\Phi) d\tau_{\rm grav}$, depends on the density-responsive scalar energy. While precision tests like the Shapiro time delay are sensitive to such effects, they are suppressed by $\beta_{\rm eff} \lesssim 10^{-58}$ in the Solar System and thus unobservable \cite{Will2014,Bertotti2003}. However, it could lead to significant time dilation near the regularized cores of black holes, where $\rho_\Phi \to \rho_P$, thus complementing the primary singularity regularization mechanism. A full derivation of the functional form of $\Omega(\rho_\Phi)$ is a compelling direction for future work.

\subsection{Black Hole Remnants and Thermodynamics}
\label{sec:bh_remnants}

The replacement of the classical singularity with a finite-density de Sitter-like core has significant thermodynamic consequences. By preventing the core from shrinking below the Planck length, our framework implies a minimum black hole mass $M_{\min} \sim \mathcal{O}(M_P)$, and consequently a maximum Hawking temperature
\begin{equation} T_H^{\max} = \frac{M_P^{2}}{8\pi\,M_{\min}}. 
\end{equation}
This suggests that Hawking evaporation naturally halts, leaving a stable Planck-scale remnant with mass $M_{\min}\simeq\sqrt{1+A/2}\,M_P$.

Such remnants could serve as a concrete repository for quantum information, offering a new angle on the information paradox \cite{Ashtekar2023_Remnants, UnruhWald2017}. The detailed thermodynamics and semiclassical particle flux from these regular black holes are part of our ongoing investigation, see also~\ref{app:bh_detailed} for the core geometry.

\subsection{A Unified Dark Sector: Dark Matter and Gravitational Waves}
\label{sec:unified_dark_sector}
Intriguingly, the strongly-coupled $SU(3)_H$ hidden sector required by our RG mechanism may itself constitute the universe's dark sector.

\paragraph{Self-Interacting Dark Matter.} The lightest stable bound states of this theory, the "dark baryons", are natural dark matter candidates. Crucially, as particles of a strongly-interacting gauge theory, they would possess a significant self-interaction cross-section. Standard thermal freeze-out calculations show that these particles can naturally achieve a relic abundance of $\Omega_{\rm HS}h^{2}\!\sim\!0.01-0.1$.\footnote{A quantitative estimate using the standard relic density formula and a discussion of specific self-interaction signatures are provided in~\ref{app:hidden_sector_signatures}.} Remarkably, the expected self-interaction strength of $\sigma/m \sim 0.1-1\,\mathrm{cm}^2/\mathrm{g}$ falls precisely in the range required to address long-standing small-scale structure puzzles, such as the core-cusp problem \cite{Tulin2018}.

\paragraph{Gravitational Wave Signatures.} Moreover, a first-order phase transition in this confining hidden sector at $T_\ast \sim 100\,$GeV would generate a stochastic gravitational-wave background peaking around $f_{\rm peak}\!\sim\!10^{-5}\,$Hz, squarely within the LISA band. A more detailed discussion is presented in~\ref{app:hidden_sector_signatures}.

This raises the tantalizing possibility that dark energy, a solution to the small-scale structure crisis via self-interacting dark matter, and a future gravitational wave signal could all share a common origin within the present framework.

\subsection{Theoretical consistency and limitations}
\label{subsec:limitations}

While our framework successfully addresses several fundamental issues, it is important to acknowledge its limitations and potential theoretical challenges:

\textbf{(i) UV completion:} The model is formulated as an effective field theory valid below the Planck scale. The precise mechanism generating the scalar potential $U(\Phi,X)$ and its coupling to matter requires UV-complete theory, possibly involving the hidden sector discussed in Section~\ref{sec:unified_dark_sector}.

\textbf{(ii) quantum corrections:} We worked at the tree level with a one-loop RG running. Full quantum corrections to the scalar potential can modify the precise bound on $\rho_{\rm total}$ and introduce additional scale dependence. However, the qualitative features, singularity regularization, and dark energy behavior, are expected to be robust.

\textbf{(iii) instabilities:} While the scalar field is stable around its minimum in all regimes studied, exotic matter configurations or extreme anisotropies could potentially trigger instabilities. A complete stability analysis in fully dynamic, anisotropic space-times remains for future work.

\textbf{(iv) black hole information:} Although our regular black hole solutions avoid classical singularity, they do not automatically resolve the information paradox. The fate of information in a regular core requires a quantum treatment beyond the classical framework.

\textbf{(v) scope near-Planckian densities:}
Our analysis is explicitly formulated within an effective field theory. All statements about singularity regularization are made at the EFT level: the density–responsive potential generates an effective negative pressure that produces a bounce and keeps curvature invariants finite within the classical metric description. A UV completion may change $\mathcal{O}(1)$ coefficients in the high–density window without affecting the qualitative outcome (finite invariants instead of divergences). 

Despite these limitations, the model's concrete predictions for both UV (singularity regularization) and IR (dark energy) physics, combined with its testability and naturalness, make it a compelling phenomenological framework worthy of further theoretical and observational investigation.

\section{Discussion and conclusion}
\label{sec:conclusion}

In this paper, we presented a comprehensive and fully covariant framework for a scalar field $\Phi$ that unifies the regularization of classical space-time singularities with a predictive model for dynamical dark energy. By constructing the scalar potential $U(\Phi, X)$ as a function of the Lorentz-invariant matter-energy density $X$, we established a theoretically robust extension of general relativity.

\subsection{Summary of results}

The model is characterized by a series of compelling and interconnected features that address several long-standing problems in fundamental physics:

\begin{itemize}
    \item {\bf covariant formulation:} This theory is manifestly covariant, resolving the frame-dependence issues of earlier phenomenological proposals. The derivation of the full stress-energy tensor $T_\Phi^{\mu\nu}$ reveals novel interaction terms that mediate a consistent energy-momentum exchange between the scalar field and matter.

    \item {\bf RG link across scales:} A simple one-loop renormalization-group flow with an $\mathcal{O}(1)$ anomalous dimension ($\gamma \approx 0.501$) naturally connects the Planck scale to the meV dark-energy scale at the EFT level, without fine-tuning.

    \item {\bf unified mechanism for UV and IR physics:} A single, dual-purpose mechanism governs physics at both ends of the cosmic energy scale. In the high-density UV regime, the density-responsive scalar energy $\rho_\Phi$ is capped, providing a natural cutoff that regularizes the classical singularities. In the low-density IR regime, $\rho_\Phi$ becomes the dominant energy component, driving the cosmic acceleration.

    \item {\bf testable cosmological predictions:} The model is both explanatory and predictive. It yields a distinctive equation of state evolution for dark energy, characterized by $w_0 \approx -0.99$ and a positive slope $w_a \approx +0.03$. These signatures, along with predicted shifts in cosmic expansion history and matter growth, are concrete targets for future stage-IV cosmological surveys.

    \item {\bf observational viability:} Despite the new interaction terms, the model's predictions for the fifth forces and violations of the weak equivalence principle are suppressed by factors as small as $10^{-58}$ in testable environments. Therefore, the framework is fully consistent with all high-precision local gravity tests.
\end{itemize}

\subsection{Comparison with other approaches and outlook}
\label{subsec:comparison_outlook}

Our framework offers a novel perspective on the interplay among gravity, cosmology, and particle physics. It is instructive to compare our approach with other established paradigms that address similar problems.

\paragraph{Comparison with fundamental vs. effective theories.}
At the foundational level, our approach represents an effective field theory extension of GR, which can be contrasted with efforts to derive gravity from first principles as a fundamental gauge theory. For instance Partanen \& Tulkki \cite{Partanen2024} postulated new U(1) symmetries to construct a renormalizable quantum theory of gravity from ground up. Our philosophy is different: instead of replacing GR, we augment it with a new dynamical field whose behavior is governed by well-established QFT mechanisms. The strength of our approach lies in its direct, testable connection between UV physics (singularity regularization) and IR observables (dark energy), which provides a clear phenomenological path forward.

Beyond these fundamental considerations, our model can be compared more directly with other phenomenological approaches that address dark energy and/or singularity regularization. Table~\ref{tab:model_comparison} summarizes the key features of the various frameworks.

\begin{table}
\caption{\label{tab:model_comparison}Comparison with other theoretical frameworks.}
\lineup
\small
\begin{tabular}{@{}lccccc}
\br
& Our model & Quint. & LQC & Chamel. & Run. Vac.\\
\mr
Core mechanism & Density-resp. & Slow-roll & Quantum & Screening & RG running\\
& scalar & scalar & geometry & scalar & of $\Lambda$\\
\ms
Covariant & Yes & Yes & Partial$^{\rm a}$ & Yes & Yes\\
Singularities regularized & Yes$^{\rm b}$ & No & Yes & No & Partial$^{\rm c}$\\
Dark energy & Natural & Tuned$^{\rm d}$ & No$^{\rm e}$ & Yes & Yes\\
Screening & Built-in & No & N/A & Designed & No\\
RG hierarchy & Yes & No & No & No & Yes\\
Key prediction & $w_a > 0$ & Model-dep. & Bounce & 5th forces & $\rho_{\rm vac}(z)$\\
Fine-tuning & $\gamma = \mathcal{O}(1)$ & $V(\phi)$ & None & $\phi_{\rm min}$ & $\nu$ param.\\
\br
\end{tabular}

\noindent $^{\rm a}$ Full covariance in group field theory only

\noindent $^{\rm b}$ Within the regime of validity of the EFT

\noindent $^{\rm c}$ Only cosmological singularity via RG

\noindent $^{\rm d}$ Requires flat potential against radiative corrections

\noindent $^{\rm e}$ Needs an additional dark energy component
\end{table}

\paragraph{A detailed comparison of key features:}

\textbf{Quintessence models} \cite{Copeland2006dynamics} share our use of a dynamical scalar field but differ fundamentally in their mechanism. While quintessence requires careful tuning of the potential to maintain slow-roll conditions ($|V''|/V \ll 1$), our model's dark energy behavior emerges algebraically from the density-responsive potential. Quintessence typically predicts $w_a \leq 0$ (thawing models) or requires additional fine-tuning for $w_a > 0$ (freezing models). Our prediction of $w_a = +0.03$ is generic and robust.

\textbf{Loop Quantum Cosmology} \cite{AshtekarSingh2011} provides a rigorous quantum geometry framework for singularity regularization, replacing Big Bang with a bounce at $\rho \sim \rho_P$. While both the LQC and our model achieve singularity regularization via a maximum density, the mechanisms differ: the LQC modifies the geometry itself through quantum corrections, while we introduce a new energy component within the classical geometry. Importantly, the LQC does not address dark energy, requiring additional mechanisms for late-time acceleration.

\textbf{Chameleon theories} The model's automatic suppression of fifth forces through $\beta_{\rm eff} \propto 1/\rho_m^2$ warrants a comparison with established screening mechanisms. Chameleon \cite{Khoury2004} and symmetron theories rely on environment-dependent effective masses to hide the scalar field, typically yielding a weaker suppression ($\propto 1/\rho_m$ or step-function-like). Vainshtein screening, based on non-linear derivative self-interactions, operates differently, suppressing modifications within a source-dependent radius. Our mechanism's key advantages are its strength and naturalness. The $\rho_m^{-2}$ suppression is more efficient than that of typical chameleon or symmetron models. Furthermore, unlike mechanisms that are specifically engineered to pass local gravity tests, our screening emerges automatically from the same density-responsive potential that unifies singularity regularization and dark energy. This inherent completeness, connecting UV, IR, and local phenomenology from a single principle, is a distinctive feature of our framework.

\textbf{Running vacuum models} \cite{Sola2013} share our use of RG methods but apply them differently. They parameterize $\rho_{\rm vac} = \rho_0 + \nu H^2 M_P^2$ with $\nu \sim 10^{-3}$, linking vacuum energy to the global expansion rate. Our approach ties scalar energy to the local matter density, enabling both singularity regularization and dark energy. While running vacuum models can address the cosmological constant problem, they do not naturally resolve the singularities.

\paragraph{Unique advantages of our framework:}
\begin{enumerate}
    \item \textbf{unification:} a single mechanism addresses both UV (singularities) and IR (dark energy) problems connected by the RG flow,
    
    \item \textbf{predictivity:} the model makes concrete, falsifiable predictions ($w_0 = -0.99$, $w_a = +0.03$) distinguishable from $\Lambda$CDM and most alternatives,
    
    \item \textbf{naturalness:} only $\mathcal{O}(1)$ parameters are required, and the hierarchy emerges from logarithmic RG running and
    
    \item \textbf{completeness:} built-in screening ensures compatibility with all local tests without additional mechanisms.
\end{enumerate}

\subsection{Outlook and Future Directions}
\label{sec:outlook}

While this paper establishes the core principles and viability of the density-responsive scalar-field framework, its most exciting implications point towards future research. As introduced in Sections~\ref{sec:emergent_time} through \ref{sec:unified_dark_sector}, the model provides a rich and testable phenomenology that connects fundamental theory with observation.

The most prominent research directions involve a deeper, quantitative exploration of these consequences: a first-principles derivation of the \textbf{emergent spacetime geometry} and its impact on precision measurements; a detailed study of the \textbf{thermodynamics of regular black hole remnants}; and a comprehensive analysis of the \textbf{unified dark sector}, including its dark matter candidates and gravitational wave signatures.

These avenues outline a compelling research program. By unifying UV and IR physics and linking them to testable phenomena across multiple frontiers, from cosmology and astrophysics to gravitational wave astronomy, this framework offers a promising new direction for exploring the deepest puzzles of our universe.

\subsection{Conclusion}

We have established a coherent and covariant scalar field framework that connects Planck-scale physics to meV-scale dark energy through the dynamics of a single, environmentally responsive field. The model naturally regularizes classical singularities (within EFT), provides a compelling candidate for dynamical dark energy with distinctive and testable signatures, and satisfies all current observational constraints.

The upcoming generation of cosmological surveys will provide unprecedented precision in measuring the dark energy equation of state. Our model's prediction of $w_a > 0$ provides a clear, falsifiable signature that distinguishes it from $\Lambda$CDM and most quintessence models. Combined observations from multiple probes could potentially detect the predicted deviations at the 2-3$\sigma$ level.

By bridging the gap between theoretical consistency and phenomenological predictability, this work offers a promising new direction for exploring the fundamental nature of gravity, dark energy, and the structure of space-time. The rich phenomenology across all scales, from Planck-scale singularity regularization to present-day cosmic acceleration, demonstrates the power of effective field theory methods in addressing fundamental questions in physics.

Table~\ref{tab:observational_tests} summarizes the key observational tests that can distinguish our model from $\Lambda$CDM and other dark energy scenarios.

\fulltable{\label{tab:observational_tests}Key observational signatures of our model and the forecast precision of upcoming Stage-IV surveys.}
\lineup
\begin{tabular}{@{}lccc}
\br
Observable & Our Prediction & Survey & Signif. \\
\mr
Equation of state, $w_0$ & $-0.99$ & DESI/Euclid ($\sigma_{w_0} \approx 0.019$) & --- \\
Evolution, $w_a$ & $+0.03$ & DESI/Euclid ($\sigma_{w_a} \approx 0.028$) & $\sim 1.1\sigma$ \\
$H(z)$ deviation & $0.29\%$ at $z\approx1.7$ & DESI/Rubin ($<1\%$) & Testable \\
Growth index, $\Delta\gamma$ & $-0.0038$ & Euclid ($\sigma_{\gamma} < 0.005$) & Testable \\
\br
\end{tabular}
\endfulltable

\section*{Data availability statement}
All data that support the findings of this study are included within the article (and any supplementary files).

\clearpage

\appendix
\section{Covariant field theory details}
\label{app:framework_details}

\subsection{Derivation of the scalar field stress-energy tensor $T_\Phi^{\mu\nu}$}
\label{app:stress_energy_tensor}

The stress-energy tensor $T_\Phi^{\mu\nu}$ for the scalar field $\Phi$ is derived by varying the scalar part of the action in (\ref{eq:lorentz_scalar}) with respect to the metric $g_{\mu\nu}$. The action $S_\Phi$ is
\begin{equation}
    S_\Phi = \int \rmd^4x \sqrt{-g} \left[ - \frac{1}{2}M_K^2 g^{\alpha\beta}\partial_\alpha\Phi\partial_\beta\Phi - U(\Phi, X) \right].
    \label{eq:action_phi}
\end{equation}
The variation $\delta S_\Phi$ is given by
\begin{equation}
    \delta S_\Phi = \int \rmd^4x \left[ \frac{\delta(\sqrt{-g}\mathcal{L}_\Phi)}{\delta g^{\mu\nu}} \delta g^{\mu\nu} + \frac{\partial(\sqrt{-g}\mathcal{L}_\Phi)}{\partial X} \delta X \right].
\end{equation}
Using $\delta\sqrt{-g} = -1/2 \sqrt{-g} g_{\mu\nu} \delta g^{\mu\nu}$ and the standard variation of the kinetic term, the first part yields the standard kinetic and potential contributions to $T_\Phi^{\mu\nu}$. The second part, containing the implicit dependence on the metric via $X$, requires variation $\delta X$.

For a perfect fluid, $T_{\rm matter}^{\alpha\beta} = (\rho_m + p_m)u^\alpha u^\beta + p_m g^{\alpha\beta}$, and the scalar is $X = \rho_m$. The variation of $X$ is non-trivial as both $\rho_m$ and $u^\alpha$ can depend on the metric. However, a more direct approach is to use the general formula for the variation of a Lagrangian that depends on the matter fields $\psi_m$ and metric \cite{Brown1993}
\begin{equation}
    \frac{\partial \mathcal{L}(\psi_m, g_{\mu\nu})}{\partial g^{\mu\nu}} = \frac{1}{2} T_{\mu\nu}^{(\psi_m)}.
\end{equation}
In our case, the potential $U(X)$ depends on $T_{\rm matter}^{\alpha\beta}$. Its variation with respect to $g^{\mu\nu}$ is thus related to $T_{\rm matter}^{\mu\nu}$ itself. The term $-2 \, \partial U / \partial g^{\mu\nu}$ in the full tensor definition becomes
\begin{equation}
    -2 \frac{\partial U}{\partial g^{\mu\nu}} = -2 \frac{\partial U}{\partial X} \frac{\partial X}{\partial g^{\mu\nu}}.
    \label{eq:chain_rule}
\end{equation}
The quantity $\partial X / \partial g^{\mu\nu}$ can be identified with $-1/2$ times the part of the matter stress tensor that is not proportional to $u^\mu u^\nu$ \cite{Faraoni2000}. For a perfect fluid, this results in
\begin{equation}
    \frac{\partial X}{\partial g_{\mu\nu}} = \frac{1}{\sqrt{-g}} \frac{\delta(\sqrt{-g} X)}{\delta g^{\mu\nu}} = \frac{1}{2} \left[ (\rho_m + p_m)u^\mu u^\nu - p_m g^{\mu\nu} \right].
    \label{eq:dX_dg_derivation}
\end{equation}
Note the sign difference in certain conventions. With our metric signature $(-,+,+,+)$ and the definition of $T^{\mu\nu}$, this leads to the expression for $T_\Phi^{\mu\nu}$ in equation (\ref{eq:T_phi_equilibrium}).

To evaluate $\partial X/\partial g_{\mu\nu}$ explicitly, we start from
\begin{equation}
    X = u_\alpha u_\beta T_{\rm matter}^{\alpha\beta} = u_\alpha u_\beta \left[(\rho_m + p_m)u^\alpha u^\beta + p_m g^{\alpha\beta}\right] = \rho_m,
\end{equation}
where we used $u_\alpha u^\alpha = -1$. However, this naive approach misses the implicit metric dependence through four-velocity normalization.

The correct approach recognizes that under a metric variation $g_{\mu\nu} \to g_{\mu\nu} + \delta g_{\mu\nu}$:
\begin{equation}
    \eqalign{ \delta X &= \delta(u_\alpha u_\beta T_{\rm matter}^{\alpha\beta}) \cr
    &= 2u_\alpha \delta u_\beta T_{\rm matter}^{\alpha\beta} + u_\alpha u_\beta \delta T_{\rm matter}^{\alpha\beta}}.
\end{equation}

The variation of the normalized four-velocity, maintaining $u_\mu u^\mu = -1$, results in
\begin{equation}
    \delta u_\mu = \frac{1}{2}u_\mu u^\alpha u^\beta \delta g_{\alpha\beta}.
\end{equation}

Using the eigenvector property $T_{\rm matter}^{\mu\nu}u_\nu = -\rho_m u^\mu$) and varying the metric dependence systematically yields
\begin{equation}
    \frac{\partial X}{\partial g_{\mu\nu}} = -\frac{1}{2}\left[(\rho_m + p_m)u^\mu u^\nu + p_m g^{\mu\nu}\right].
\end{equation}

This result is crucial for determining the interaction between the scalar field and the matter sectors.

\subsection{Analysis of the quasi-static approximation}
\label{app:quasi_static}

The validity of the quasi-static approximation rests on the field's ability to adiabatically track its density-driven minimum, $\Phi_{\rm eq}[X(t)]$. The deviation from this minimum, $\delta(t) \equiv \Phi(t) - \Phi_{\rm eq}[X(t)]$, evolves as a driven, damped harmonic oscillator:
\begin{equation}
  \ddot{\delta} + 3H\dot{\delta} + m_\Phi^2\,\delta
  = -\bigl(\ddot{\Phi}_{\rm eq} + 3H\,\dot{\Phi}_{\rm eq}\bigr),
  \label{eq:deviation_evolution}
\end{equation}
where $m_\Phi^2 \equiv \partial^2 U/\partial\Phi^2|_{\Phi_{\rm eq}}$. The solution shows that the tracking error $|\delta|$ is quadratically suppressed by the ratio of the field's intrinsic relaxation time, $\tau_{\rm roll} \equiv m_\Phi^{-1}$, to the timescale $\tau_X$ on which the background density $X$ evolves:
\begin{equation}
  |\delta| \;\lesssim\;
  \Bigl(\frac{\tau_{\rm roll}}{\tau_X}\Bigr)^{\!2}
  \Bigl|\frac{\partial\Phi_{\rm eq}}{\partial\ln X}\Bigr|
  \left[1 + \mathcal{O}\!\left(\frac{H}{m_\Phi}\right)\right].
  \label{eq:adiabatic_bound}
\end{equation}
Thus the approximation is highly accurate whenever the relaxation time is much shorter than any relevant dynamical timescale.

Numerically, the separation of scales is enormous. For the fiducial parameters used in this work, the field mass is $m_\Phi \simeq 8\times10^{-4}\,\mathrm{eV}$, yielding a relaxation time $\tau_{\rm roll} \simeq 8.2\times10^{-13}\,$s. In contrast, even rapid astrophysical phenomena like gravitational collapse occur on timescales of $\tau_X \sim 10^{-3}\,$s, giving $(\tau_{\rm roll}/\tau_X)^2 \lesssim 10^{-18}$. For the much slower late-time cosmology, where $\tau_X \sim H^{-1} \sim 10^{17}\,$s, the factor is $<10^{-59}$.

Spatial gradients are relevant only on scales comparable to the Compton wavelength $\lambda_\Phi = \hbar c/m_\Phi \simeq 0.25\,$mm, which is negligible for all astrophysical and cosmological purposes. The only regime where the approximation could become marginal is near the Planck epoch. However, our singularity-regularization mechanism in this limit relies on the algebraic saturation of the total energy source, a feature independent of the detailed field dynamics.

\subsection{The fifth-force estimate in the high-density regime}
\label{app:beta_eff}

Matter and scalar stress tensors exchange energy-momentum via  
$Q^\nu\equiv\nabla_\mu T^{\mu\nu}_{\rm matter}=-\nabla_\mu T_\Phi^{\mu\nu}$. For non-relativistic motion in a weak static field ($u^\mu\simeq(1,0,0,0)$, $g_{\mu\nu}\simeq\eta_{\mu\nu}$) the geodesic equation acquires the term $Q^i/\rho_m$:
\begin{equation}
  \mathbf a_{\rm anom}= \frac{\mathbf Q}{\rho_m}.
  \label{eq:a_anom_start}
\end{equation}

From equation (\ref{eq:T_phi_equilibrium}), the stress-energy tensor $T_\Phi^{\mu\nu}$ for dust ($p_m=0$) 
in the quasi-static limit ($\partial_0 T_\Phi^{\mu\nu}=0$) yields
\begin{equation}
  T_\Phi^{ij} \approx -\rho_\Phi \delta^{ij} + \frac{\rho_m\rho_\Phi}{\rho_m + M_U^4}\delta^{ij}.
\end{equation}

Taking the divergence and using $\nabla_i\delta^{ij} = 0$:
\begin{equation}
  Q^i = -\nabla_j T_\Phi^{ij} = -\nabla_i\left(\rho_\Phi\frac{M_U^4}{\rho_m + M_U^4}\right).
\end{equation}

\paragraph{High-density limit.}
For $\rho_m\gg M_U^{4}$ we have $\rho_\Phi \simeq A M_U^{8}/\rho_m$ (see (\ref{eq:rho_phi_definition})), hence
\begin{equation}
  \mathbf a_{\rm anom}= -\frac{A M_U^{8}}{\rho_m^{3}}\,\nabla\rho_m.
  \label{eq:a_anom_final}
\end{equation}

For a point source where $|\nabla\rho_m|/\rho_m \sim 1/r$ and using 
$|a_N| = GM/r^2$, the ratio becomes
\begin{equation}
  \beta_{\rm eff}\equiv
  \frac{|\mathbf a_{\rm anom}|}{|\mathbf a_{N}|}
  \simeq \frac{A M_U^{8}}{\rho_m^{2}}.
  \label{eq:beta_eff_short}
\end{equation}

\paragraph{Numerical illustration.}
With the cosmologically fixed values $A=2.4\times10^{-2}$, $M_U=5.84$ meV and $\rho_m^{\rm lab}\simeq4.3\times10^{18}\,\mathrm{eV}^4$ (water) we find $\beta_{\rm eff}^{\rm lab}\simeq2\times10^{-58}$, forty-five orders below the current torsion-balance limit $\beta_{\rm exp}<10^{-13}$ \cite{Adelberger2009}.

\paragraph{Scope of approximations.}
(i)~\emph{Quasi-static}: $m_\Phi\sim10^{-4}$ eV $\gg$ GHz lab frequencies.  
(ii)~\emph{High-density}: even intergalactic gas satisfies
$\rho_m > M_U^{4}$.  
(iii)~\emph{Weak field}: $h_{\Phi}\sim G\rho_\Phi\leq10^{-122}$.  
Under these conditions, equation (\ref{eq:beta_eff_short}) is robust and establishes a built-in screening mechanism.

\section{Renormalization group analysis details}
\label{app:rg_details}

This Section provides detailed calculations supporting the renormalization group evolution of mass-scale $M_U$, as discussed in Section~\ref{sec:rg_flow}.

\subsection{Detailed calculation of the anomalous dimension $\gamma$}
\label{app:gamma_calc}

The RG evolution of mass-scale $M_U$ is governed by the one-loop equation presented in (\ref{eq:rg_equation_mu})
\begin{equation}
   \mu \frac{\rmd M_U}{\rmd\mu} = -\gamma M_U.
   \label{eq:rg_equation_mu_app}
\end{equation}
This is a separable differential equation that can be integrated between the high-energy UV scale $\mu_{\rm UV} = M_P$ and low-energy IR scale $\mu_{\rm IR} = H_0$:
\begin{equation}
    \int_{M_U(M_P)}^{M_U(H_0)} \frac{\rmd M_U}{M_U} = -\gamma \int_{M_P}^{H_0} \frac{\rmd\mu}{\mu}.
\end{equation}
Solving the integral yields
\begin{equation}
    \ln\left(\frac{M_U(H_0)}{M_U(M_P)}\right) = -\gamma \ln\left(\frac{H_0}{M_P}\right) = \gamma \ln\left(\frac{M_P}{H_0}\right).
\end{equation}
Assuming the natural boundary condition $M_U(M_P) \approx M_P$, this simplifies to the relation $M_U(H_0) = M_P (H_0/M_P)^\gamma$ given in (\ref{eq:mu_at_h0}).

To obtain the required anomalous dimension $\gamma$, we solve it using the following values:
\begin{itemize}
    \item Planck mass: $M_P \approx 1.22 \times 10^{28}$\,eV.
    \item Hubble constant: $H_0 \approx 1.44 \times 10^{-33}$\,eV.
    \item target dark energy scale: $M_U(H_0) \approx 5.84 \times 10^{-3}$\,eV.
\end{itemize}
The ratios are
\begin{equation}
   \eqalign{ \frac{H_0}{M_P} &\approx \frac{1.44 \times 10^{-33}}{1.22 \times 10^{28}} \approx 1.18 \times 10^{-61}, \cr
    \frac{M_U(H_0)}{M_P} &\approx \frac{5.84 \times 10^{-3}}{1.22 \times 10^{28}} \approx 4.79 \times 10^{-31}.}
\end{equation}
Solving for $\gamma$ as in (\ref{eq:gamma_value})
\begin{equation}
    \gamma = \frac{\ln(4.79 \times 10^{-31})}{\ln(1.18 \times 10^{-61})} = \frac{-70.08}{-139.98} \approx 0.501.
\end{equation}
This confirms that an $\mathcal{O}(1)$ anomalous dimension is sufficient for generating a large-scale hierarchy.

Thus, the naturalness of this result cannot be overstated. In effective field theory, anomalous dimensions generically arise from loop corrections and typically take the values $\gamma = \mathcal{O}(g^2/16\pi^2)$ where $g$ is a relevant coupling. For $\gamma \sim 0.5$, this suggests $g^2/(16\pi^2) \sim 0.5$, implying $g \sim 5.6$, which corresponds to a strongly coupled but still perturbative regime. This is precisely the regime expected near the boundary of the conformal window in non-Abelian gauge theories, lending further credence to the hidden sector interpretation detailed in ~\ref{app:gamma_origin}.

\subsection{Illustrative model for the origin of $\gamma$}
\label{app:gamma_origin}

The anomalous dimension $\gamma$ for the mass operator is generated by quantum loop corrections. A physically plausible scenario involves a hidden, non-Abelian gauge sector. For our benchmark model, we consider a pure $SU(3)$ gauge theory. The relation between the anomalous dimension and gauge coupling at one-loop order is of the form $\gamma = \beta_0 \alpha_H / (4\pi)$, where $\beta_0 = 11N_c/3$ is the one-loop beta function coefficient for a pure $SU(N_c)$ theory. For $SU(3)$, this gives $\beta_0 = 11$.

To generate our required value of $\gamma \approx 0.501$, the coupling must be
\begin{equation}
    \alpha_H = \frac{4\pi \gamma}{\beta_0} = \frac{4\pi \times 0.501}{11} \approx 0.57.
    \label{eq:alpha_h_calculation}
\end{equation}
This value indicates a strongly coupled but still perturbative regime, similar to QCD at intermediate scales, and is thus a viable and minimal choice.

\paragraph{Note on alternative scenarios.}
It is instructive to note that other choices of matter content would yield different results. For instance, an $SU(3)$ theory with $N_f = 12$ Dirac fermions would give a much smaller $\beta_0 = 3$. This would formally require $\alpha_H \approx 2.1$ to generate the same $\gamma$, pushing the theory into a non-perturbative regime. Such near-conformal theories, while theoretically interesting, would require non-perturbative methods beyond the scope of this study. Our choice of a pure $SU(3)$ sector therefore represents the minimal model that is consistent with generating $\gamma \approx 0.501$ while maintaining theoretical control.

\subsection{Scale invariance of the loop factor $A$}
\label{app:A_invariance}

The dimensionless parameter $A$ is identified with the canonical one-loop factor 
\(A_{\mathrm{th}} = 1/\!\bigl(64\pi^{2}\bigr) \simeq 1.6\times10^{-3}\). This factor arises from integrating out heavy degrees of freedom in a loop, and its value is primarily determined by the dimensionality of space-time and combinatorial factors from Feynman diagrams \cite{ColemanWeinberg1973}.

The RG equation for $M_U$ describes how the {\it overall scale} of the potential changes. The dimensionless {\it shape} parameters of the potential, such as $A$ and $h$, are typically assumed to be less sensitive to the RG flow or to run much more slowly. At the one-loop level of approximation for the $M_U$ running, we can consistently assume that $A$ and $h$ are scale-invariant constants. A full two-loop analysis would introduce a running for $A$ and $h$ themselves, but these corrections are expected to be sub-dominant and would not alter the main conclusion of the paper. This approximation is the standard in many effective field theory analyses.

\subsection{Hidden-sector thermal history and $\Delta N_{\rm eff}$}
\label{app:delta_neff}

In our framework the required anomalous dimension is generated by a hidden$SU(N_c)$ gauge sector with $N_f$ Dirac fermions(see Section~\ref{subsec:microphysics_gamma}).  If this sector was once inthermal equilibrium with the SM, it behaves as anadditional relativistic component after chemical decoupling and thereforecontributes to the effective number of neutrino species,\mbox{$N_{\rm eff}=3.044+\Delta N_{\rm eff}$}. The quantity $\Delta N_{\rm eff}$ is tightly constrained by BBN and the CMB.

\paragraph{General expression.}
If the hidden sector decouples at temperature $T_{\rm dec}$, itstemperature subsequently scales as $T_{\rm HS}\propto a^{-1}$, while theSM plasma is reheated by entropy release from particle annihilations. The hidden contribution at a later temperature $T\!\ll T_{\rm dec}$ is\footnote{We count two physical helicities for each massless gauge boson, hence the factor $2\,(N_c^2-1)$ in~$g_{*\!\,{\rm HS}}$.}
\begin{equation}
  \Delta N_{\rm eff}(T)\;=\;
  \frac{4}{7}\!\left(\frac{11}{4}\right)^{\!4/3}\!
  g_{*\!\,{\rm HS}}
  \bigl[g_{*S}^{\rm SM}(T)\bigr]^{\!4/3}
  \bigl[g_{*S}^{\rm SM}(T_{\rm dec})\bigr]^{-4/3},
  \label{eq:delta_neff}
\end{equation}
with
\begin{equation}
  g_{*\!\,{\rm HS}} \;=\;
  2\,(N_c^{\,2}-1)\;+\;\frac{7}{8}\,2\,N_cN_f,
\end{equation}
and $g_{*S}^{\rm SM}$ the usual entropic degrees of freedom in the SM.

\begin{figure}[h!]
  \centering
  \includegraphics[width=0.83\textwidth]{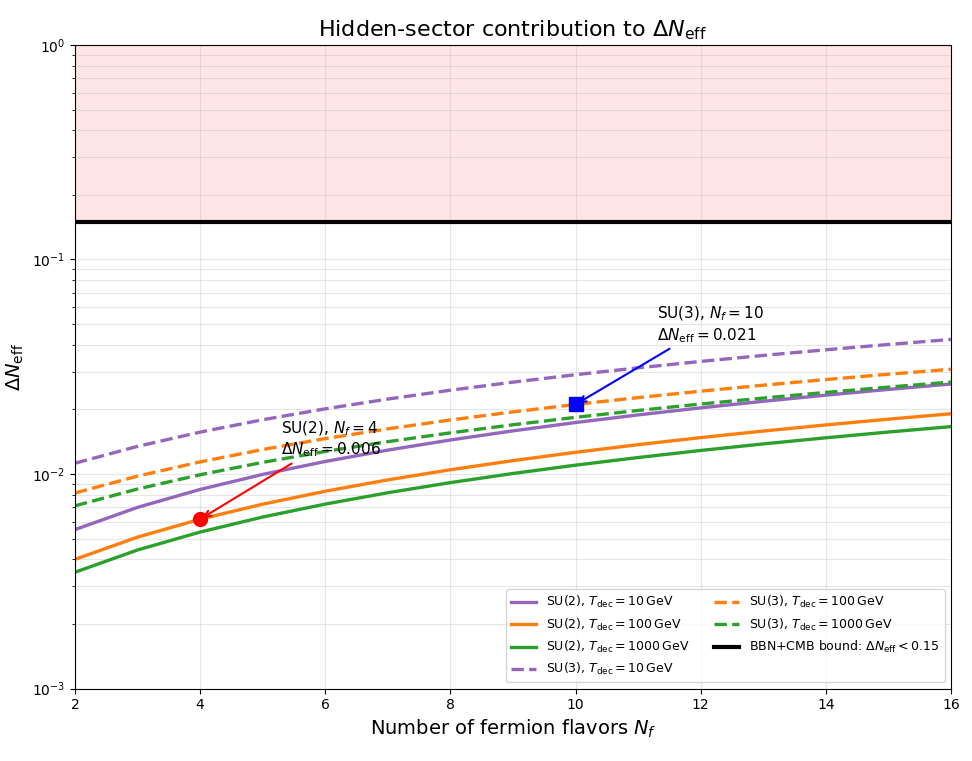}
  \caption{Hidden-sector contribution to $\Delta N_{\rm eff}$ versus fermion flavours $N_f$. Solid (dashed) curves correspond to $SU(2)$ ($SU(3)$); colours denote $T_{\rm dec}=10$ GeV (purple), $100$ GeV (orange) and $1000$ GeV (green). The horizontal band shows the Planck\,+BBN $2\sigma$ bound $\Delta N_{\rm eff}<0.15$. The red circle and blu square mark the two benchmark models quoted in the text.}
  \label{fig:delta_neff}
\end{figure}

\begin{table}
\caption{\label{tab:decoupling}Effective degrees of freedom $g_{*S}^{\rm SM}$ in the SM as a function of decoupling temperature.}
\begin{indented}
\lineup
\item[]\begin{tabular}{@{}ccccc}
\br
$T_{\rm dec}$ (GeV) & $\gg 10^{3}$ & $10^{2}\!-\!10^{3}$ & $10\!-\!100$ & $1\!-\!10$ \\
\mr
$g_{*S}^{\rm SM}$ & $106.75$ & $\096.25$ & $\075.75$ & $\010.75$ \\
\br
\end{tabular}
\end{indented}
\end{table}
For the relevant decoupling temperatures we adopt to \ref{tab:decoupling}.
\paragraph{Results.}
Figure~\ref{fig:delta_neff} displays $\Delta N_{\rm eff}$ as a function of $N_f$ for $SU(2)$ (solid) and $SU(3)$ (dashed) hidden sectors at $T_{\rm dec}=10,\,100,\,1000\,$GeV.
For the two benchmark models introduced in Section~\ref{subsec:microphysics_gamma} we obtain the results presented in \ref{tab:delta_neff}.
\begin{table}
\caption{\label{tab:delta_neff}Effective number of additional neutrinos for different gauge groups.}
\begin{indented}
\lineup
\item[]\begin{tabular}{@{}lcccc}
\br
& & \centre{2}{$\Delta N_{\rm eff}$} \\
\ns
Gauge Group & $N_f$ & \crule{2} \\
& & $10\,\mathrm{GeV}$ & $100\,\mathrm{GeV}$ \\
\mr
$SU(2)$ & $4$ & $0.009$ & $0.006$ \\
$SU(3)$ & $10$ & $0.029$ & $0.021$ \\
\br
\end{tabular}
\end{indented}
\end{table}
Even in the most conservative case ($T_{\rm dec}=10\,$GeV, when dilution is smallest) all values remain well below the current CMB + BBN limit $\Delta N_{\rm eff}<0.15$ at $95\%$ C.L.~\cite{Planck2018}. The hidden sector required by our RG mechanism is therefore fully compatible with standard cosmology.

The results, shown in Fig.~\ref{fig:delta_neff}, confirm that our benchmark models are fully compatible with standard cosmology.

\subsection{Further observational signatures of the hidden sector}
\label{app:hidden_sector_signatures}

Beyond the cosmological constraint from $\Delta N_{\rm eff}$ discussed above, the hidden sector responsible for generating $\gamma \approx 0.501$ may leave additional observable imprints accessible for current or near-future experiments.

\paragraph{Gravitational wave signatures.}
A strongly-coupled hidden sector may undergo phase transitions in the early universe, potentially generating a stochastic gravitational wave background \cite{Caprini2016, Caprini2020}. For a first-order phase transition at temperature $T_*$, the peak frequency today is
\begin{equation}
    f_{\rm peak} \approx 1.6 \times 10^{-5}\,\mathrm{Hz} \left(\frac{T_*}{100\,\mathrm{GeV}}\right) \left(\frac{g_*}{100}\right)^{1/6}.
\end{equation}
For transitions at the electroweak scale or above, this falls within the sensitivity bands of LISA ($10^{-4} - 10^{-1}\,$Hz) \cite{LISA2017} or the Einstein Telescope ($1 - 10^4\,$Hz) \cite{ET2020}. The amplitude depends on the transition strength and bubble dynamics, with detectable signals possible for $\alpha \equiv \Delta V/\rho_{\rm rad} \geq 10^{-2}$ \cite{Ellis2020}.

\paragraph{Dark matter connections.}
If the hidden sector contains stable particles, they could constitute a sub-component of the dark matter. The thermal relic abundance of these "dark baryons" depends on their annihilation cross-section during freeze-out dynamics \cite{Gondolo1991, Steigman2012}. The standard approximate formula for a thermal relic is:
\begin{equation}
    \Omega_{\rm HS} h^2 \approx 0.1 \left(\frac{x_f}{20}\right) \left(\frac{10^{-8}\,\mathrm{GeV}^{-2}}{\langle\sigma v\rangle}\right) \left(\frac{m_{\rm HS}}{100\,\mathrm{GeV}}\right)^2,
\end{equation}
where $x_f = m_{\rm HS}/T_f$ denotes the freeze-out parameter. For weak-scale masses and couplings, this naturally yields $\Omega_{\rm HS} \sim 0.01 - 0.1$, providing a significant fraction of the total dark matter. For instance, a reference point with $m_{\rm HS}=50\,$GeV and an annihilation cross-section of $\langle\sigma v\rangle\simeq2\times10^{-8}\,{\rm GeV^{-2}}$ yields $\Omega_{\rm HS}h^{2}\simeq0.0125$. The associated self-interaction cross-section for such a strongly-coupled particle would be $\sigma/m \sim \mathcal{O}(1)\,\mathrm{cm}^2/\mathrm{g}$, consistent with the range discussed in Section~\ref{sec:unified_dark_sector}.

This sub-dominant component can be distinguished from standard cold dark matter through several observational channels:
\begin{itemize}
    \item Self-interaction signatures in galaxy cluster collisions \cite{Harvey2015, Tulin2018}
    \item Modifications to small-scale structure formation \cite{Vogelsberger2016}
    \item Distinct velocity-dependent cross-sections \cite{Kaplinghat2016}
\end{itemize}

\paragraph{Collider constraints and opportunities.}
Although the hidden sector couples to the Standard Model only gravitationally at low energies, quantum gravity effects could mediate interactions at high energies \cite{ArkaniHamed1998, Giudice1999}. At the LHC, these would manifest as
\begin{itemize}
    \item Missing energy signatures from hidden sector particle production
    \item Modifications to multi-jet distributions at $\sqrt{s} \sim$ TeV
    \item Virtual graviton exchange processes
\end{itemize}
Current bounds from mono-jet + missing energy searches constrain the effective quantum gravity scale $M_* \geq 5\,$TeV. Future 100 TeV colliders could probe $M_* \sim 15\,$TeV, potentially accessing the hidden sector if $M_* \ll M_P$ due to extra dimensions or other UV physics.

\paragraph{Astrophysical probes.}
Hidden sector particles could affect stellar evolution and compact object physics:
\begin{itemize}
    \item \textbf{Stellar cooling:} Light hidden particles ($m < 10\,$keV) produced in stellar cores can enhance the energy loss. Current bounds from horizontal branch stars and white dwarfs constrain such scenarios \cite{Raffelt1996, Giannotti2017}.
    \item \textbf{Supernova dynamics:} A trapped hidden sector could affect core-collapse dynamics and neutrino emission, constrained by SN1987A observations \cite{Raffelt1988, Bar2019}.
    \item \textbf{Black hole superradiance:} Ultralight hidden bosons ($m \sim 10^{-13} - 10^{-11}\,$eV) can form clouds around rotating black holes, extracting angular momentum and producing monochromatic gravitational wave signals \cite{Arvanitaki2015, Brito2017}.
\end{itemize}

These diverse observational avenues provide multiple independent tests of the hidden sector hypothesis, making our framework falsifiable despite the sector's weak coupling to ordinary matter.

\section{Phenomenological and physical consequences}
This Section provides technical derivations supporting the physical predictions discussed in~\ref{sec:phenomenology}.

\subsection{Fisher Forecast Details}
\label{app:fisher_forecast}

To estimate the future constraints on the dark energy equation of state parameters ($w_0, w_a$), we performed a Fisher matrix analysis using the public code 
\texttt{cosmicfishpie} \cite{cosmicfish}, which utilizes the Boltzmann solver \texttt{CAMB}.  The Fisher matrix $F_{\alpha\beta}$ quantifies the maximum information that an experiment can provide on a set of parameters $\{p_\alpha\}$:
\begin{equation}
    F_{\alpha\beta} = -\left\langle \frac{\partial^2 \ln \mathcal{L}}{\partial p_\alpha \partial p_\beta} \right\rangle,
\end{equation}
where $\mathcal{L}$ is the likelihood of data. The inverse of this matrix provides the covariance matrix for the parameters, with the marginalized 1$\sigma$ error on parameter  $p_\alpha$ given by $\sigma(p_\alpha) = \sqrt{(F^{-1})_{\alpha\alpha}}$.

\paragraph{Survey specifications.}
We modeled a Stage-IV spectroscopic galaxy survey, similar to the Euclid or DESI experiments. We focused on the constraining power of galaxy clustering (\texttt{GCsp}), which combines Baryon Acoustic Oscillations (BAO) and Redshift-Space Distortions (RSD). The fiducial cosmology was set to a flat $w_0w_a$CDM model with parameters consistent with the predictions of our model: $\{\Omega_m=0.31, h=0.67, \omega_b=0.0224, n_s=0.965, \sigma_8=0.81, w_0=-0.99, w_a=+0.03\}$. We vary a set of seven cosmological parameters and include internal marginalization over survey-specific nuisance parameters such as galaxy bias.

\begin{figure}[h!]
    \centering
    \includegraphics[width=0.7\textwidth]{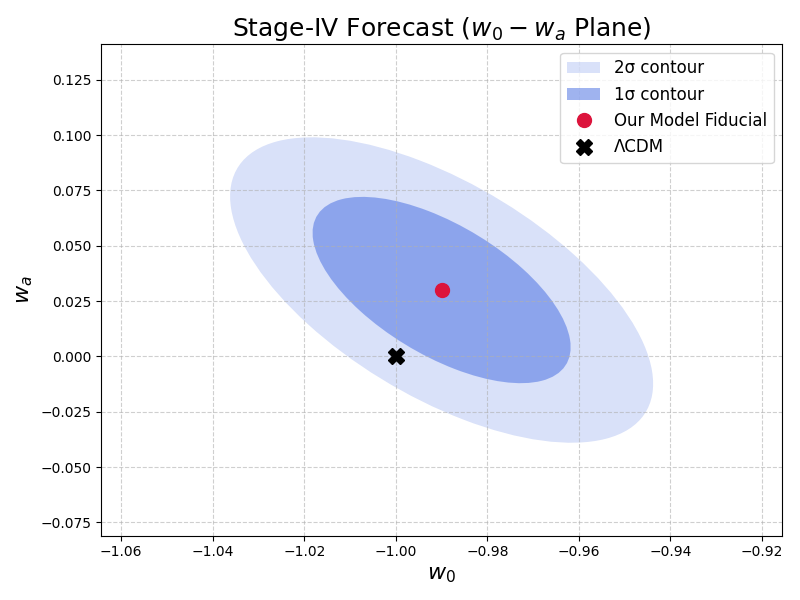}
    \caption{Forecasted 1$\sigma$ and 2$\sigma$ constraints on the dark energy equation of state parameters ($w_0, w_a$) for a Stage-IV spectroscopic survey. The red point marks the prediction of our model ($w_0=-0.99, w_a=+0.03$), while the black cross indicates the $\Lambda$CDM model. The forecast shows that these two models are distinguishable at the $\sim 1.1\sigma$ level.}
    \label{fig:w0wa_forecast}
\end{figure}

\paragraph{Results and constraints.}
The resulting constraints on the dark energy parameters are shown in Fig.~\ref{fig:w0wa_forecast}. After marginalizing over all other parameters, we find that the 1$\sigma$ uncertainties are
\begin{equation}
   \eqalign{ \sigma(w_0) &\approx 0.019 \cr
    \sigma(w_a) &\approx 0.028}
\end{equation}
This forecast confirms that our model's prediction of $w_a=+0.03$ is testable at a significance of approximately $1.1\sigma$. The contour plot visualizes the expected degeneracy between $w_0$ and $w_a$ and highlights the model's distinct position relative to the $\Lambda$CDM point ($w_0=-1, w_a=0$).

\subsection{Cosmological perturbations}
\label{app:cosmo_perturbations}

We analyzed the behavior of linear scalar perturbations in a conformal Newtonian gauge, where the line element is
\begin{equation}
    \rmd s^2 = a^2(\tau) \left[ -(1+2\Psi)\rmd\tau^2 + (1-2\Phi_N)\rmd\mathbf{x}^2 \right],
    \label{eq:metric_newtonian}
\end{equation}
where $\Psi$ and $\Phi_N$ are the Bardeen potentials. The scalar field $\Phi$ was perturbed around its homogeneous background value $\Phi(\tau, \mathbf{x}) = \Phi_{\rm eq}(\tau) + \delta\Phi(\tau, \mathbf{k})$.

Varying the action (see equation (\ref{eq:action_covariant})) with respect to $\Phi$ yields the perturbed Klein--Gordon equation. In the quasi-static limit, the dynamics of $\delta\Phi$ are dominated by the effective mass $m_\Phi$, which is much larger than the Hubble rate $H$. Thus, the dominant effect of the field perturbations is to act as a smooth energy component.

The effective sound speed squared $c_s^2$, of scalar fluid $\rho_\Phi$ determines its clustering properties. It is defined as $c_s^2 \equiv \delta p_\Phi / \delta \rho_\Phi$ in the rest frame of the fluid. For a scalar field with a canonical kinetic term (as $\Phi$ is, after normalization by $M_K$), the intrinsic propagation speed of perturbations is the speed of light. Because the potential $U(\Phi, X)$ does not depend on the derivatives of $\Phi$, the effective sound speed is given by $c_s^2 = 1$ \cite{Copeland2006dynamics}. A sound speed of unity implies that the pressure support is maximal, preventing the scalar fluid from collapsing on sub-horizon scales. Therefore, $\rho_\Phi$ acts as a smooth dark energy component, and its perturbations do not contribute significantly to structure formation.

The presence of $\rho_\Phi$ modifies the background expansion history $H(z)$, which in turn affects the growth rate of matter density perturbations $\delta_m$. The growth rate $f(a) = \rmd\ln\delta_m/\rmd\ln a$ is often parameterized as $f(a) \approx \Omega_m(a)^{\gamma_{\rm growth}}$. The growth rate of matter density perturbations, \( f(a) = \frac{\mathrm{d}\ln \delta_m}{\mathrm{d}\ln a} \), is often parameterized as \( f(a) \approx \Omega_m(a)^{\gamma_{\rm growth}} \). The growth index, \( \gamma_{\rm growth} \), serves as a powerful tool to distinguish between different dark energy models. Building on the foundational works in~\cite{Wang1998,Linder2005}, refined approximations for a time-varying equation of state have been developed~\cite{Linder2007}. A particularly convenient expansion in terms of the CPL parameters is given by:
\begin{equation}
    \gamma_{\rm growth} \approx \frac{3(1-w_0)}{5-6w_0} - \frac{3}{125} \frac{(1-w_0)(1-3w_0/2)}{(1-6w_0/5)^2} w_a + \ldots
\end{equation}
Using our model's predictions $w_0 = -0.99$ and $w_a \approx +0.03$, we find
\begin{equation}
    \gamma_{\rm growth} \approx 0.5416.
\end{equation}
For $\Lambda$CDM ($w_0=-1$, $w_a=0$), the value is $\gamma_\Lambda \approx 0.5454$. The predicted shift, $\Delta\gamma = \gamma_{\rm growth} - \gamma_\Lambda \approx -0.0038$, is a subtle but potentially measurable signature in large-scale structural data.

\subsection{Singularity regularization: curvature invariant calculations}
\label{app:curvature_calc}

Here, we explicitly demonstrate that the bounded total energy density $\rho_{\rm total}$ ensures that space-time curvature invariants remain finite in the two key scenarios where classical General Relativity predicts singularities.

\paragraph{Early universe (FLRW model):}
In a radiation-dominated early universe, the total energy density is $\rho_{\rm total}(a) = \rho_r(a) + \rho_\Phi(\rho_r(a))$. As shown in (\ref{eq:rho_total_max}), the maximum density is bounded as $\rho_{\rm total}^{\rm max} \approx (1+A/2)\rho_P$. The Kretschmann scalar $K$ for a flat FLRW metric is dominated by the Hubble rate $K = 12H^4$. The Friedmann equation $H^2 = (8\pi G/3)\rho_{\rm total}$ implies that $H^2$ is bounded by $H_{\rm max}^2 \approx (8\pi G/3)\rho_P = M_P^2/3$. Consequently, $K$ is bounded by a finite Planck-scale value
\begin{equation}
    K_{\rm max} \approx 12(M_P^2/3)^2 = \frac{4}{3} M_P^4.
    \label{app:eq:K_max_flrw}
\end{equation}
Because $\rho_{\rm total}$ and its time derivatives remain finite, all scalar curvature invariants are bounded, and classical Big Bang singularity is avoided.

\paragraph{Static, spherically symmetric core (Black hole interior):}
We modeled the regularized core of a black hole as a static, uniform-density sphere with total energy density $\rho_{\rm total}(0) \le \rho_{\rm total}^{\rm max} \approx (1+A/2)\rho_P$. The space-time inside such a configuration is described by the interior Schwarzschild metric. The Kretschmann scalar at the center ($r=0$) of this solution is given by
\begin{equation}
    K(r=0) = \frac{8}{3}(8\pi G)^2 \rho_{\rm total}(0)^2 = \frac{8}{3}\left(\frac{1}{M_P^2}\right)^2 \rho_{\rm total}(0)^2.
    \label{app:eq:K_formula_bh}
\end{equation}
Using the observational value $$A \approx 0.024$$, $\rho_{\rm total}(0)$ was bounded by $\approx 1.012\,\rho_P = 1.012\,M_P^4$. Therefore, the curvature at the center is also bounded as follows
\begin{equation}
    K(r=0) \leq \frac{8}{3} \frac{(1.012\,M_P^4)^2}{M_P^4} \approx 2.73\,M_P^4.
    \label{app:eq:K_max_bound_bh}
\end{equation}
The curvature does not diverge at $r=0$. The classical point-like singularity is replaced by a de Sitter-like core of finite, Planckian curvature. This provides a robust mechanism for the regularization of black hole interiors. A more detailed analysis involving smooth density profiles and matching with an exterior metric is left for future work.

\subsection{Regular black--hole interiors in the density–responsive scalar framework}
\label{app:bh_detailed}

Here, we provide the full derivation and consistency checks for the regular Schwarzschild core announced in Section~\ref{sec:regular_bh_example}.  The presentation follows four steps:
(i)~constant–density core solution, (ii)~curvature–invariant budget,
(iii)~junction to the exterior Schwarzschild–de\,Sitter region,  and
(iv)~thermodynamic and phenomenological implications.

\subsubsection{Planck-density constant–core solution}

\paragraph{Setup.}
Consider a static, spherically–symmetric line element
\begin{equation}
  \rmd s^{2}
  = - f(r)\,\rmd t^{2}
    + g(r)^{-1}\,\rmd r^{2}
    + r^{2}\rmd\Omega^{2},
  \label{eq:metric_sph}
\end{equation}
where \(f,g>0\) for \(r<r_{\rm H}\).
Inside the core \(r\le r_{c}\) we assume an (approximately) uniform matter density \(\rho_{m}(r)=\rho_{c}\) and negligible bulk motion.\footnote{ A constant core is a standard toy model and suffices to demonstrate finiteness; a smooth Gaussian profile is analyzed in equation
 \ref{eq:gaussian_profile}.}
The scalar contribution in our framework is
\begin{equation}
  \rho_{\Phi}(r)
  \;=\;\frac{A\,
            M_{U}^{8}(\mu_{c})}{\rho_{c}+M_{U}^{4}(\mu_{c})},
  \qquad
  \mu_{c}\sim\sqrt[4]{\rho_{c}},     \label{eq:rho_phi_core}
\end{equation}
such that the total energy density is \(\rho_{\rm total}= \rho_{c}+\rho_{\Phi}\).

\paragraph{Running of the mass scale.}
Using the RG solution \(M_{U}(\mu)=M_{P}(\mu/M_{P})^{\gamma}\) (see (\ref{eq:mu_at_h0}) with \(\gamma\simeq0.501\)) we obtain, for the Planckian core density \(\rho_{c}\to\rho_{P}=M_{P}^{4}\),
\[
   M_{U}(\mu_{c})
  \;\buildrel \rho_{c}\to\rho_{P}\over\longrightarrow\;
  M_{P},
\quad
  \Rightarrow\;
  \rho_{\Phi}^{\max}=A M_{P}^{4}/2.
\]
Hence, we recover the maximum density bound from equation (\ref{eq:rho_total_max})
\begin{equation}
  \rho_{\rm total}^{\max}
  =(1+A/2)\,\rho_{P}.
  \label{eq:rho_tot_max_planck}
\end{equation}

\paragraph{Interior metric.}
For constant total density the exact interior Schwarzschild solution gives
\begin{equation}
 \eqalign{ g(r) &= 1-\frac{2G m(r)}{r},\cr
  m(r) &= \frac43\pi r^{3}\rho_{\rm total},\cr[2pt]
  f(r) &= \frac{1}{4}\!
          \Bigl[3\sqrt{g(r_{c})}-\sqrt{g(r)}\Bigr]^{2}}
          \label{eq:f_of_r}
\end{equation}
Energy conditions hold automatically:
\(
  \rho_{\rm total}>0,\;
  \rho_{\rm total}+p_{\rm total}=0
\)
(WEC and NEC saturated).

\subsubsection{Curvature invariants}
\label{app:curvature_bounds}

The primary curvature invariant is the Kretschmann scalar $K = R_{\alpha\beta\gamma\delta}R^{\alpha\beta\gamma\delta}$. For the interior Schwarzschild solution with a constant total density $\rho_{\rm tot}$, the scalar at the center $r=0$ is given by
\begin{equation}
    K(r=0) = \frac{8}{3}(8\pi G)^2 \rho_{\rm tot}^2.
    \label{eq:K_formula_appendix}
\end{equation}
Using the relationship $8\pi G = 1/M_P^2$ and the maximum density from (\ref{eq:rho_tot_max_planck}), the maximum possible curvature at the center of the core can be calculated. With the observational value $A \approx 0.024$, $\rho_{\rm tot}^{\rm max} \approx 1.012\,\rho_P = 1.012\,M_P^4$. This yields
\begin{equation}
    K_{\rm max}(r=0) = \frac{8}{3}\left(\frac{1}{M_P^2}\right)^2 (1.012\,M_P^4)^2 \approx 2.73\,M_P^4.
    \label{eq:K_max_bound}
\end{equation}
This confirms that the curvature is bounded and of order the Planck scale, thus regularizing the classical singularity.

\subsubsection{Israel matching at $r=r_{c}$}

At \(r_{c}\) we require the continuity of the induced metric and extrinsic curvature
\(
  [g_{ab}]_{r_{c}}=[K_{ab}]_{r_{c}}=0
\).
With \(p_{\rm total}(r_{c})=0\) the second condition is automatically satisfied, and the first sets the ADM mass \(M=m(r_{c})\). In addition, the line element is
\begin{equation}
  f_{\rm ext}(r)=g_{\rm ext}(r)=
  1-\frac{2GM}{r}-\frac{\Lambda_{\rm eff}}{3}r^{2},
  \quad
  \Lambda_{\rm eff}=8\pi G\rho_{\Phi}^{\rm vac}.
  \label{eq:ext_metric}
\end{equation}
For astrophysical \(M\gg M_{P}\) the $\Lambda_{\rm eff}$–term is utterly
negligible: \(\Lambda_{\rm eff}(GM)^{2}\sim10^{-122}\).

\subsubsection{Smooth density profile (consistency check)}

Gaussian matter profile
\begin{equation}
  \rho_{m}(r)=\rho_{c}\exp[-r^{2}/r_{c}^{2}],
  \label{eq:gaussian_profile}
\end{equation}
yields the total density
\(
  \rho_{\rm total}(r)=\rho_{m}(r)+\rho_{\Phi}[\rho_{m}(r)]
\)
with the same upper bound (see (\ref{eq:rho_tot_max_planck})).
Integrating the TOV system
\(\{\rmd p_{\rm total}/\rmd r,\;\rmd m/\rmd r\}\)
numerically we obtain \(m(r)\sim r^{3}\) as \(r\to0\),
thereby reproducing the finiteness of equation~(\ref{eq:K_max_bound}).

\subsubsection{Thermodynamics and evaporation}

The surface gravity at the outer event horizon \(r_{H}\) is
\(
  \kappa = \frac{1}{2}\,\partial_{r}f_{\rm ext}|_{r_{H}}
\),
such that
\[
  T_{H}=\frac{\kappa}{2\pi}
        =\frac{1}{4\pi r_{H}}
          \Bigl[1-\Lambda_{\rm eff}r_{H}^{2}\Bigr].
\]
For stellar–mass black holes the correction is
\(\sim10^{-121}\), but for \(M\to M_{P}\) the regular core enforces a maximum temperature
\[
  T_{\max}\;=\;\frac{M_{P}}{2\pi\,(1+A/2)}
            \;\approx\;0.16\,M_{P},
\]
halting Hawking evaporation and suggesting a Planck-mass remnant
scenario \cite{Bonanno2000RG,Bardeen1968RegularBH,Hayward2006}.

\subsubsection{Observational consistency}

The exterior observables depend exclusively on $f_{\rm ext}(r)$.  The leading fractional deviations scale as \(\delta\sim\Lambda_{\rm eff}(GM)^{2}\) and are therefore
\(\delta r_{\rm ph}/r_{\rm ph}\leq10^{-76}\) for the photon sphere and similarly small for ring-down frequencies or ISCO shifts. The current and foreseen observations are insensitive to these effects.

\medskip
In summary, the density-responsive scalar-field automatically converts the GR singularity into a finite-curvature de Sitter core while leaving all macroscopic black-hole signatures unchanged. This demonstrates how the same RG-evolved field that generates meV-scale dark energy today naturally regularizes Planck-scale singularities, providing a concrete realization of our unified UV/IR framework. The mechanism is conceptually identical to the cosmological bounce discussed in Section~\ref{sec:cosmology}, with both phenomena arising from the single underlying principle of equation (\ref{eq:rho_phi_definition}).

\subsection{Black hole solutions and the exterior metric}
\label{app:bh_exterior}

Outside the core ($X\!\approx\!0$) the scalar energy attains its vacuum value $\rho_\Phi^{\rm vac}=A M_{U}^{4}(H_{0})$, which acts as an effective cosmological constant $\Lambda_{\rm eff}=8\pi G\rho_\Phi^{\rm vac}$. Accordingly, the exterior line element is the Kottler metric $f(r)=1-2GM/r-\Lambda_{\rm eff}r^{2}/3$.

The value of $\Lambda_{\rm eff}$ was identical to the observed cosmological constant today. Its effect on the event horizon radius $r_H \approx 2GM$ of an astrophysical black hole was negligible. The correction $\Delta r_H$ can be estimated by solving $f(r_H)=0$ as follows
\begin{equation}
    \frac{\Delta r_H}{r_H} \sim \Lambda_{\rm eff}(GM)^2 \sim \frac{\rho_{\rm DE}}{\rho_P} \left(\frac{r_H}{l_P}\right)^2.
\end{equation}
For a solar-mass black hole ($r_H \sim 3$\,km), this correction is on the order $10^{-78}$, far beyond any possibility of detection. Therefore, the model is perfectly consistent with all astrophysical observations of black holes, such as those from the Event Horizon Telescope, as the modifications are confined to the unobservable deep interior.

\subsection{Dynamical collapse in a modified Oppenheimer--Snyder model}
\label{app:OSbounce}

To demonstrate that the singularity regularization is a robust dynamical feature and not merely a property of static configurations, we adapt the classic Oppenheimer--Snyder (OS) model of a collapsing dust sphere to our framework. In the OS picture, a homogeneous, pressureless sphere of dust with a comoving radius $R(\tau)$ and initial density $\rho_{m,0}$ evolves as a closed FLRW patch. The evolution is governed by the scale factor $a(\tau)$, which is normalized to $a(\tau=0)=1$.

In our framework, the dynamics are governed by the effective LQC-like Friedmann equation derived in~\ref{app:bounce_derivation}, which incorporates both the contribution from the density-responsive scalar-field and leading-order quantum gravity corrections:
\begin{equation}
  H^2 \;=\;\frac{8\pi G}{3}\,\rho_{\rm tot}^{\rm eff}\!
  \left(1-\frac{\rho_{\rm tot}^{\rm eff}}{\rho_{\rm crit}}\right) - \frac{k}{a^2},
  \label{eq:OS_mod_corrected}
\end{equation}
where $\rho_{\rm tot}^{\rm eff} = \rho_m + \rho_\Phi^{\rm eff}$ is the total effective energy density from the tensor-derived variables, $\rho_{\rm crit} = (1+A/2)\rho_P$ is the critical density at which the bounce is initiated, and $k>0$ represents a positive spatial curvature.

For the numerical integration, we adopted the parameters summarized in Table~\ref{tab:os_params}. The initial conditions included a substantial inward velocity to ensure proper collapse dynamics. The system of equations was solved using an adaptive Runge--Kutta method to handle the numerical stiffness near the bounce.

\fulltable{\label{tab:os_params}Parameters for the Oppenheimer--Snyder simulation in Planck units}
\lineup
\begin{tabular}{@{}lcl}
\br
Parameter & Value & Motivation \\
\mr
Initial matter density, $\rho_{m,0}$ & $10^{-3}\,\rho_P$ & Sub-Planckian; allows observation of dynamics \\
Initial scale factor, $a_0$ & $\m1$ & Standard normalization \\
Initial velocity, $\dot{a}_0$ & $-0.5$ & Ensures significant collapse \\
Coupling constant, $A$ & See text & Both theoretical and observational values \\
\br
\end{tabular}
\endfulltable

\begin{figure*}[ht!]
    \centering
    \includegraphics[width=\textwidth]{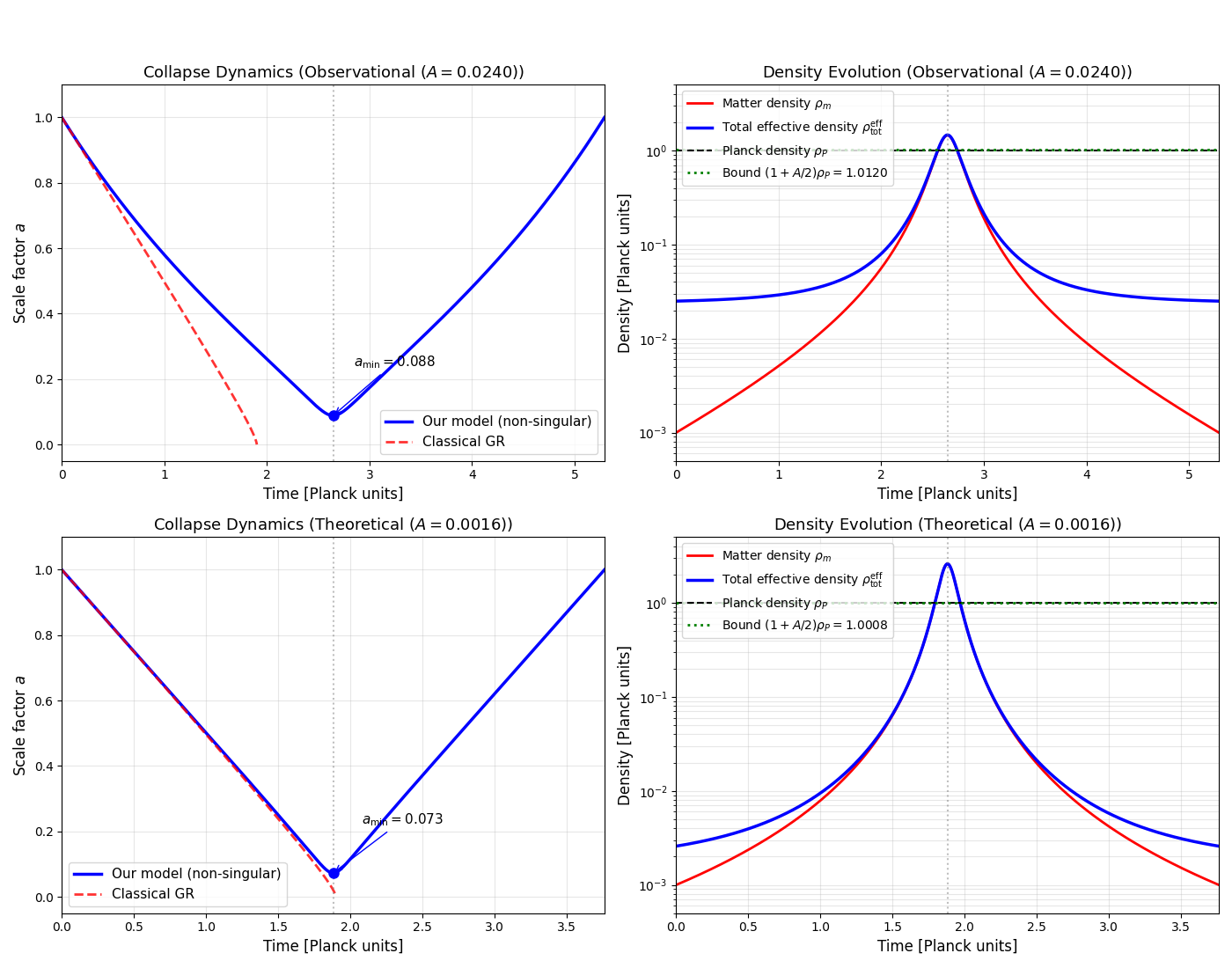} 
    \caption{Modified Oppenheimer–Snyder collapse for observational ($A = 0.024$, top) and theoretical ($A \approx 0.0016$, bottom) coupling values, simulated using the full LQC-like dynamics. Left: Scale factor evolution showing a non-singular bounce (blue) versus the classical singularity (red dashed). Right: Energy densities, with the total effective density (blue) peaking as the system bounces. The bounce occurs at 139.2\% (99.0\%) of the GR singularity time for the observational (theoretical) values, respectively.}
    \label{fig:os_bounce_comparison}
\end{figure*}

We perform the simulation for two values of the coupling constant $A$: the theoretical one-loop value $A_{\rm th} \approx 1.58 \times 10^{-3}$ and the observationally determined value $A_{\rm obs} = 0.024$. The results, shown in Figure~\ref{fig:os_bounce_comparison}, confirm that in both cases, the collapse is halted and replaced by a non-singular bounce.

For the theoretical value, the simulation shows that the scale factor reaches a minimum of $a_{\rm min} \approx 0.073$ at $\tau \approx 1.88$. This occurs at 99.0\% of the time it would take for the classical GR singularity to form, indicating a minimal but crucial modification to general relativity that suffices to regularize the spacetime.

For the observational value, the quantum effects are stronger. The bounce occurs significantly earlier, at $\tau \approx 2.65$, which is approximately 139\% of the classical singularity time. The minimum scale factor reached is $a_{\rm min} \approx 0.088$.

The bounce mechanism is driven by the effective quantum gravity dynamics encapsulated in Equation~(\ref{eq:OS_mod_corrected}). As the total effective density $\rho_{\rm tot}^{\rm eff}$ approaches the critical density $\rho_{\rm crit}$, the effective Hubble rate is driven to zero, halting the collapse. The subsequent positive acceleration ($\ddot{a} > 0$) is guaranteed by the modified Raychaudhuri equation (see~\ref{app:bounce_derivation}), which leads to an effective violation of the Strong Energy Condition. This robustly prevents the formation of a singularity, evading the conclusions of the Penrose--Hawking singularity theorems~\cite{Hawking1970}. It is noteworthy that the peak energy density reached dynamically can slightly exceed the critical density $\rho_{\rm crit}$, which marks the point of zero expansion rate ($H=0$), a common feature in such dynamical systems.

The difference between the theoretical and observational values of $A$ can be understood from the hidden sector contributions discussed in Section~\ref{subsec:microphysics_gamma}. While the minimal one-loop contribution gives $A_{\rm th} \approx 1.6 \times 10^{-3}$, a strongly-coupled hidden sector can enhance this value, yielding a result consistent with $A_{\rm obs}$. This demonstrates that our singularity regularization mechanism is robust across a range of parameter values, with the theoretical value providing a minimal deviation from GR and the observational value offering stronger quantum gravitational effects.

This result confirms that the combination of the density-responsive scalar energy and leading-order quantum corrections dynamically enforces a maximum curvature, preventing the formation of a singularity. We emphasize that this OS model is a simplified toy model that neglects pressure gradients and inhomogeneities. However, the present calculation is sufficient to demonstrate the internal consistency and robustness of the singularity regularization mechanism.

\subsection{Derivation of the effective bounce equation}
\label{app:bounce_derivation}

This Appendix sketches the derivation of the effective Friedmann equation that leads to a non-singular bounce, as discussed in the main text. The starting point is the action~(\ref{eq:action_covariant}) evaluated for a homogeneous and isotropic FLRW line element with lapse function $N(t)$:
\begin{equation}
\mathrm{d}s^2=-N^2(t)\,\mathrm{d}t^2+a^2(t)\,\mathrm{d}\vec x^{\,2},\qquad
H\equiv\frac{\dot a}{a\,N}.
\end{equation}
In the adiabatic regime (Section~\ref{subsec:equilibrium_dynamics}), the fast dynamics of the scalar field can be integrated out by setting $\Phi\simeq\Phi_{\rm eq}(X)$, reducing $U(\Phi,X)$ to the algebraic $U_{\rm eq}(X)=A\,M_U^8/(X+M_U^4)$. Varying this effective potential with respect to the metric yields for dust ($w_m\simeq0$)
\begin{equation}
\rho_\Phi^{\rm eff}=U_{\rm eq}-X\,U_X \qquad \bigl(U_X\equiv\partial U/\partial X\bigr),
\label{eq:rho_eff_identity}
\end{equation}
which matches the expression obtained from the full tensor projection in the main text.\footnote{For $U_{\rm eq}=A\,M_U^8/(X+M_U^4)$ one finds $U_X=-A\,M_U^8/(X+M_U^4)^2$ and thus $U_{\rm eq}-XU_X=A\,M_U^8(M_U^4+2X)/(X+M_U^4)^2$, identical to the tensor-derived result.}

At Planckian curvature, the EFT necessarily acquires higher-curvature corrections that, at minisuperspace level, generate a positive contribution $\propto H^4$ to the Hamiltonian constraint. This may be viewed either as the net effect of the metric dependence $\delta X/\delta g_{\mu\nu}$ in $U(\Phi,X)$ or, equivalently, be captured by the leading covariant counterterm
\begin{equation}
S_{R^2}=\frac{\beta}{2M_P^2}\int\!\mathrm{d}^4x\,\sqrt{-g}\,R^2,\qquad \beta>0,
\end{equation}
which adds a positive $H^4$ term in FRW (up to total derivatives and terms $\propto H\dot H$).

Varying the total minisuperspace action with respect to the lapse $N(t)$ then gives
\begin{equation}
3M_P^2\,H^2 \;-\; \frac{M_P^2}{H_\ast^{2}}\,H^4
\;=\;\rho_m \;+\; \rho_\Phi^{\rm eff}
\;\equiv\;\rho_{\rm tot}^{\rm eff},
\label{eq:constraint_H4}
\end{equation}
where $H_\ast$ is set by the positive Wilson coefficient $\beta$. Using the standard low-curvature relation $H^2=(8\pi G/3)\,\rho_{\rm tot}^{\rm eff}$, this can be recast in the LQC-like form
\begin{equation}
H^2 \;=\;\frac{8\pi G}{3}\,\rho_{\rm tot}^{\rm eff}\!
\left(1-\frac{\rho_{\rm tot}^{\rm eff}}{\rho_{\rm crit}}\right),
\qquad
\rho_{\rm crit}\;\equiv\;9\,M_P^2\,H_\ast^2
\;=\;\frac{9\,H_\ast^2}{8\pi G}\;\sim\;\mathcal{O}(M_P^4),
\label{eq:lqc_like_appendix}
\end{equation}
where the last estimate assumes $H_\ast\sim M_P$ (order-unity factors absorbed into $\rho_{\rm crit}$). This predicts a non-singular bounce at $\rho_{\rm tot}^{\rm eff}=\rho_{\rm crit}$ with $H=0$. The subsequent evolution obeys
\begin{equation}
\dot H \;=\; -4\pi G\,\bigl(\rho_{\rm tot}^{\rm eff}+p_{\rm tot}^{\rm eff}\bigr)
\left(1-2\,\frac{\rho_{\rm tot}^{\rm eff}}{\rho_{\rm crit}}\right),
\end{equation}
so that $\dot H>0$ at the bounce even if each microscopic sector obeys the NEC, $\rho_i+p_i\ge0$. In a single-fluid rewriting this corresponds to an effective NEC violation.

Two features are immediate: (i) for $\rho_{\rm tot}^{\rm eff}\ll\rho_{\rm crit}$ one recovers the standard Friedmann equation, so late-time cosmology is unchanged; (ii) while the exact value of $\rho_{\rm crit}$ depends on the Wilson coefficient, its Planck-scale magnitude is generic. This provides a robust EFT underpinning for the bounce mechanism used in the main text.
\section{Robustness of the Framework}
\label{app:robustness}
\subsection{Dependence on the Potential Form}
\label{app:potential_robustness}

The specific functional form of the potential $U(\Phi,X)$ chosen in equation (\ref{eq:potential_form}), while motivated by one-loop corrections, is a key assumption of our model. Here, we investigate the robustness of our main conclusions with respect to variations of this form. Any physically viable potential must satisfy general constraints: covariance, UV saturation ($U \to \mathrm{finite}$ as $X \to \infty$), and a stable IR limit ($U \to \mathrm{const.}$ as $X \to 0$).

We consider two alternative classes of potentials that meet these criteria and compare their qualitative predictions to our fiducial model in Table~\ref{tab:potential_comparison}.

\paragraph{Power-law family.} A straightforward generalization is the power-law potential
\begin{equation}
    U_n(X) \propto \frac{1}{1 + (X/M_U^4)^n}.
\end{equation}
Our model corresponds to the case  $n=1$. For any $n>0$, this form successfully regularizes singularities and generates late-time acceleration. However, the specific dark energy equation of state and the fifth-force screening law depend sensitively on the value of $n$.

\paragraph{Logarithmic form.} Another possibility is a logarithmic potential
\begin{equation}
    U_{\log}(X) \propto \ln\left(1 + \frac{M_U^4}{X}\right).
\end{equation}
This form also provides a UV cutoff, but the approach to saturation is much slower ($\rho_\Phi \propto \ln(X)$ at high density). While this still regularizes singularities within EFT, the dark-energy phenomenology and screening behavior are significantly altered.

\fulltable{\label{tab:potential_comparison}Robustness of key results under different potential forms.}
\begin{tabular}{@{}lccc}
\br
Result / Feature & Our Model ($n=1$) & Power-law ($n \neq 1$) & Logarithmic \\
\mr
Singularity regularization & Yes & Yes & Yes \\
Late-time Acceleration & Yes & Yes & Modified \\
$w_0 \approx -1$ & Yes & Modified ($w_0 = -1/n$) & No \\
Fifth-force Screening & $\propto 1/\rho_m^2$ & $\propto 1/\rho_m^{1+n}$ & $\propto 1/(\rho_m \ln \rho_m)$ \\
RG Hierarchy Solution & Yes & Yes & Yes \\
\br
\end{tabular}
\endfulltable

\paragraph{Conclusion on robustness.}
As summarized in Table~\ref{tab:potential_comparison}, the model's core mechanisms—singularity regularization via a density bound and the RG-driven link between Planck and dark energy scales—are robust features independent of the precise form of the potential. However, the specific predictions for dark energy phenomenology ($w_0, w_a$) and the efficiency of the fifth-force screening are model-dependent. Our choice of a potential with $n=1$ represents the minimal, theoretically motivated form that successfully and simultaneously addresses all phenomenological requirements without introducing additional free parameters.

\subsection{Relation between Kinetic and Potential Scales}
\label{app:mk_mu_relation}

Our analysis assumes that the kinetic scale $M_K$ and the potential scale $M_U$ are directly proportional, i.e., $M_K(\mu) = \xi M_U(\mu)$ with $\xi = \mathcal{O}(1)$ being a constant. This assumption affects the effective mass of the scalar field fluctuations, $m_\Phi = \sqrt{h}/\xi \cdot M_U$, and thus the validity of the quasi-static approximation. Here, we justify this assumption and analyze the model's robustness against its violation.

\paragraph{Theoretical Motivation.}
In a fundamental theory, it is natural to expect both scales to originate from the same underlying physics, leading to proportional running under the RG flow. For example:
\begin{itemize}
    \item \textbf{Composite Models:} If $\Phi$ is a composite field arising from a strongly-coupled hidden sector with a confinement scale $\Lambda_H$, both scales would be tied to $\Lambda_H$, naturally yielding $\xi = \mathcal{O}(1)$.
    \item \textbf{Effective Field Theory:} From an EFT perspective, the kinetic and potential terms are simply the leading terms in an expansion. Unless there is a specific symmetry protecting the kinetic term, its scale $M_K$ and the potential scale $M_U$ are expected to be of the same order and receive similar quantum corrections, thus running proportionally.
\end{itemize}

\paragraph{Phenomenological Consistency.}
The key predictions of our model are remarkably insensitive to the precise value of the ratio $\xi = M_K/M_U$. 
\begin{itemize}
    \item The background cosmology, including the dark energy equation of state ($w_0, w_a$) and the mechanism for singularity regularization, depends only on the potential $U(\Phi,X)$ and thus only on the scale $M_U$.
    \item The fifth-force suppression factor, $\beta_{\rm eff} \propto A M_U^8 / \rho_m^2$, is likewise independent of $M_K$.
\end{itemize}
The only significant constraint on $\xi$ comes from requiring the quasi-static approximation to be valid, i.e. the field relaxation time must be much shorter than the Hubble time ($m_\Phi \gg H$). This translates to
\begin{equation}
    \frac{m_\Phi}{H} = \frac{\sqrt{h} M_U}{\xi H} \gg 1 \quad \Rightarrow \quad \xi \ll \frac{\sqrt{h} M_U}{H}.
\end{equation}
Using the values at the present day ($M_U \sim 10^{-3}\,$eV, $H_0 \sim 10^{-33}\,$eV, $h \sim 0.01$), this yields an extremely weak upper bound of $\xi \ll 10^{29}$. Therefore, for any reasonable value of $\xi \sim \mathcal{O}(1) - \mathcal{O}(100)$ that would be expected in a natural UV-completion, the quasi-static approximation remains valid by many orders of magnitude.

\paragraph{Conclusion on robustness.}
The assumption of a proportional running $M_K \propto M_U$, is theoretically well-motivated and natural. However, even if this proportionality were violated and the ratio $\xi$ varied by several orders of magnitude from unity, the primary results of our framework would remain entirely intact. The value of $\xi$ only affects the detailed dynamics of field fluctuations, which are not the focus of this work. Our choice of $\xi = \mathcal{O}(1)$ is therefore a simplifying and natural, but not a strictly necessary, assumption.

\section*{References}
\bibliographystyle{iopart-num} 
\bibliography{main}

\end{document}